\documentclass[12pt]{article}

\usepackage{graphicx,amsmath,amssymb,physics} 
\usepackage{fullpage}

\usepackage{hyperref}
\usepackage[style=verbose-ibid,backend=bibtex]{biblatex}
\graphicspath{{Images/}}
\numberwithin{equation}{section}
\title{An English Translation of Gröbli's Ph.D. Dissertation: ``Specielle Probleme über die Bewegung geradliniger paralleler Wirbelfäden''}
\author{Roy Goodman}

\date{\today}

\begin{document}

\maketitle

\section*{Note}
Walter Gröbli's 1877 doctoral dissertation was the first study of the dynamics of $n\ge3$ interacting vortices in an inviscid incompressible two-dimensional fluid~\autocite{Grobli:1877}. It casts a long shadow over the field, and his method for integrating the three-vortex problem has been applied many times.

In studying these problems, I have wanted to see for myself exactly what Gröbli knew and how his mathematical approach differed from that of modern dynamical systems.  Aref, Rott, and Thomann researched Gröbli's life and his short mathematical career and have summarized how he reduced the three-vortex problem to quadratures~\autocite{Aref.1992} but, as far as I have determined, no complete English translation exists. Despite being illiterate in German, I decided to write one. I will include my observations on the text in some upcoming publications.

This translation was made possible by software. Google has scanned the dissertation and made it available on the \href{https://www.google.com/books/edition/Spezielle_Probleme_über_die_Bewegung_ge/hgzDSAAACAAJ?hl=en}{web}. I used \href{https://mathpix.com}{Mathpix Snip} to recognize text and equations in the scanned PDF. In addition to turning printed math into \LaTeX, its AI produced meaningful German text much more reliably than the Optical Character Recognition (OCR) built into Adobe Acrobat Professional. This text was then copied into Google Translate. I then used my best judgment to make sense of its returned results.

\subsection*{Some observations and warnings}

\paragraph{Correctness} While I have spent a lot of time correcting the equations produced by the OCR, and the English and specialized mathematical grammar of Google's translation, I have not systematically compared the equations produced with those in the original dissertation, nor have I checked the correctness of most of Gröbli's computations. I suspect there are many errors, which would have been harder to find and fix than in today's computer-based publishing workflow. Caveat Lector.

\paragraph{Equation numbering} Gröbli restarts the equation numbers several times throughout the text, for reasons I haven't determined. I have chosen to use the modern \LaTeX\ convention of numbering by section. Many numbered equations are never referenced, but I defer to Gröbli's judgment here and number the equations he numbers.

\paragraph{Section names} Gröbli doesn't consistently title his sections. In addition, there are some bold headings between sections, but these are not formalized into sections or chapters following the current convention. I have left this as is.

\paragraph{Notation} I stuck to his notation with one exception I changed the alternate kappa $\varkappa$ to $\kappa$ since the former is hard to distinguish from an $x$. There were a lot of mistakes in the original where $\kappa$ was replaced by $k$ here and there. These are fixed.

\paragraph{Starting a sentence with a variable name} He does it. I don't like it. Sometimes, I rewrite the sentence, and sometimes, I don't.

\pagebreak

\includegraphics[width=\textwidth]{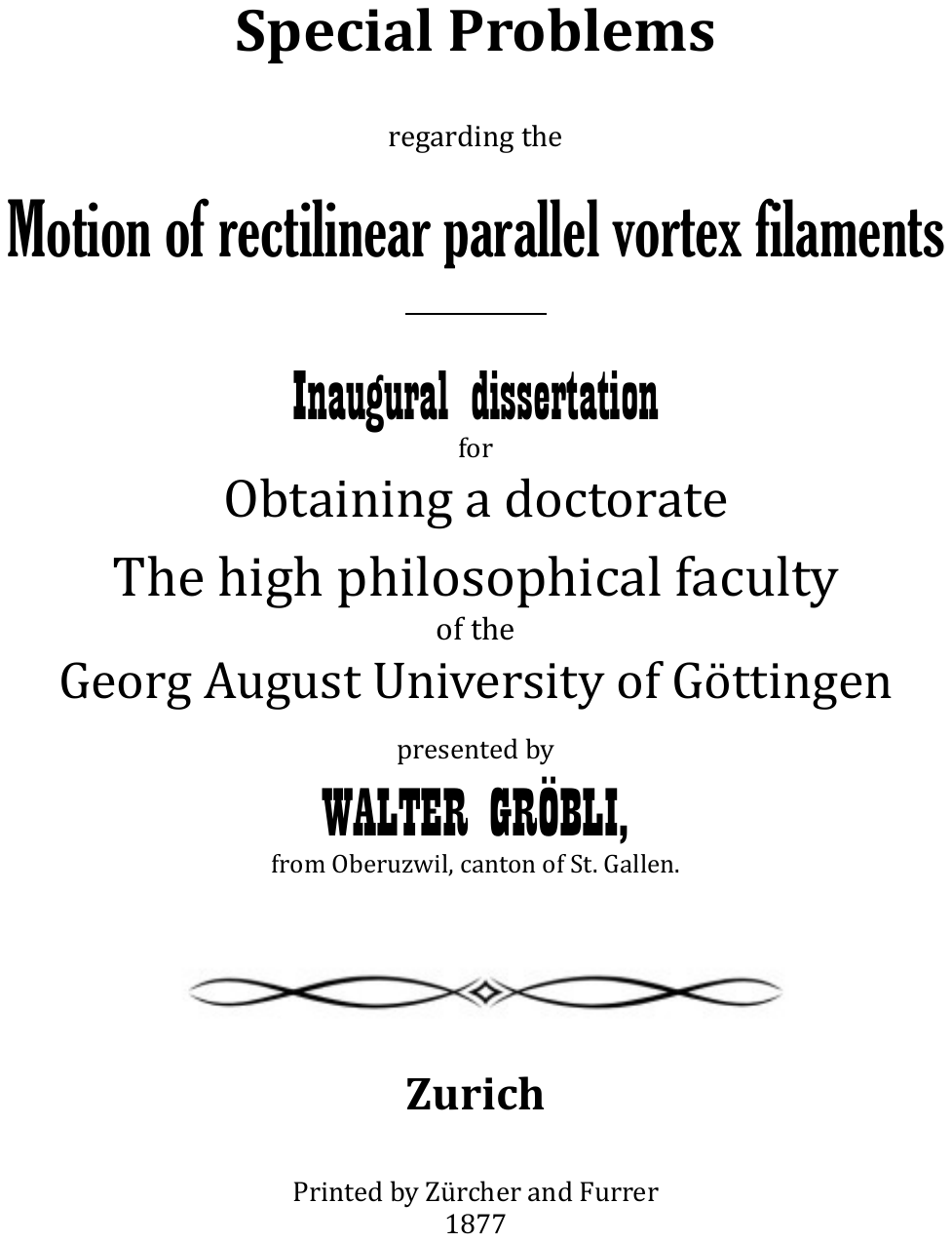}

\paragraph{Dedication} Dedicated to his loving parents, Isaac Gröbli and Elisabetha Gröbli, née Grob, from the author.
\\

The present investigations arose from the written work that I had to submit in the summer of 1875 as a student in the mathematical section of the specialist teaching department of the Federal Polytechnic to obtain my diploma. In this work, for which I received the topic from my esteemed teacher, Professor Dr. Heinrich Weber, now in Königsberg, I have reworked most of it and added several new things.

\section{}
\label{section1}

The vortex theory deals with liquid movements in which the individual liquid particles may also have rotational movements. A line whose direction coincides everywhere with the direction of the momentary axis of rotation of the water particles located there is called a vortex line, according to Helmholtz~\autocite{Helmholtz:1858}. All the vortex lines through the points of an infinitely small closed curve cut out a thread of vortices from the liquid. A whirlpool continually comprises the same fluid particles, either returning to itself or terminating at the surface of the liquid; the product of the rotational speed in the cross sections is the same for all cross sections and at all times.

In the present work, it is assumed that the liquid is bounded by two planes perpendicular to the $z$-axis, extends between them to infinity, and is stationary at infinity, that the motion is parallel to the $xy$-plane and independent of the $z$-coordinate. The vortex threads, the number of which should be finite, are then parallel to the $z$-axis. Let $\dd f$ be an element of the cross-section of one of the vortex threads with the $xy$-plane and $\zeta$ the speed of rotation of this element; we define a quantity $m$ by the equation
\begin{equation} \label{1_1}
  m = \int \zeta \dd f,
\end{equation}
where the integration is to be extended over the cross-section of this thread. Since the motion is the same in all planes parallel to the $xy$ plane, it suffices to determine it in that plane. Let the orthogonal coordinates of the vortex filaments be
$$
x_1, \, y_1; \quad x_2,\, y_2; \quad x_3, \, y_3,\ldots
$$
the values of the associated constants $m$
$$
m_1, m_2, m_3, \ldots
$$

The differential equations, which determine the movement of the vortex threads, are then according to Kirchhoff~\autocite{Kirchhoff:1876}.
\begin{equation} \label{1_2}
\begin{aligned}
m_1 \dv{x_1}{t}&=\phantom{-} \pdv{P}{y_1}, & m_2 \dv{x_2}{t}&= \phantom{-}\pdv{P}{y_2}, \ldots \\
m_1 \dv{y_1}{t}&=-\pdv{P}{x_1}, & m_2 \dv{y_2}{t}&=-\pdv{P}{x_2}, \ldots
\end{aligned}
\end{equation}
where
\begin{equation} \label{1_3}
P = - \frac{1}{\pi} \sum m_1 m_2 \log{\varrho_{12}}.
\end{equation}
The variable $\varrho_{12}$ denotes the distance between threads 1 and 2, and the sum is taken over all combinations of two different indices.

If one introduces polar coordinates $(\varrho, \vartheta)$ by means of the equations
\begin{equation} \label{1_4}
\begin{aligned}
  x_1 & = \varrho_1 \cos{\vartheta_1}, & x_2 & = \varrho_2 \cos{\vartheta_2}, \ldots \\
  y_1 & = \varrho_1 \sin{\vartheta_1}, & y_2 & = \varrho_2 \sin{\vartheta_1}, \ldots
\end{aligned}
\end{equation}
then Eq.~\eqref{1_2} becomes:
\begin{equation} \label{1_5}
\begin{aligned}
 m_1 \varrho_1\dv{\varrho_1}{t}&= \pdv{P}{\vartheta_1}, 
& m_2 \varrho_2\dv{\varrho_2}{t}&= \pdv{P}{\vartheta_2}, \ldots \\
 m_1 \varrho_1\dv{\vartheta_1}{t}&=-\pdv{P}{\varrho_1}, 
& m_2 \varrho_2\dv{\vartheta_2}{t}&=-\pdv{P}{\varrho_2}, \ldots
\end{aligned}
\end{equation}
The following four integrals are known from equations~\eqref{1_2} and~\eqref{1_5}.
\begin{equation} \label{1_6}
\begin{aligned}
  \sum m_1 x_1 & = \text{const.}, & \sum m_1 y_1 & = \text{const.}, \\
  \sum m_1 \varrho_1^2 & = \text{const}, & P &= \text{const}.
\end{aligned}  
\end{equation}
If we think of the rotational speed $\zeta$ as the density of a mass spread out on the element $\dd f$, the first two of these integrals express the theorem that the center of mass distribution, which one can call the center of gravity of the vortex threads since only the vortex threads contribute to it, remains at rest.

If only one vortex thread is present, it remains in its place; if there are two vortex filaments, they rotate with constant angular velocity
$$
\frac{1}{\pi} \frac{m_1 m_2}{m_1 \varrho_1^2 + m_2 \varrho_2^2}
$$
around the focus.

In what follows, we shall deal with the movement of three vortex threads, four vortex threads assuming a plane of symmetry, and finally, $2 n$ vortex filaments subject to $n$ planes of symmetry.

We will not discuss the determination of the motion of liquid particles a finite distance from the vortex filaments.

\section{Concerning the movement of three vortex threads}
\label{section2}

We introduce the somewhat more convenient symbols $s_1$, $s_2$, and $s_3$ for the pairwise distances $\varrho_{23}$, $\varrho_{31}$, and $\varrho_{12}$ between the three vortex threads The differential equations governing the motion of the system of three vortex filaments are in Cartesian coordinates
\begin{equation} \label{2_1}
\begin{aligned}
\pi \dv{x_1}{t}&=-m_2 \frac{y_1-y_2}{s_3^2}+m_3 \frac{y_3-y_1}{s_2^2}; \\
\pi \dv{x_2}{t}&=-m_3 \frac{y_2-y_3}{s_1^2}+m_1 \frac{y_1-y_2}{s_3^2}; \\
\pi \dv{x_3}{t}&=-m_1 \frac{y_3-y_1}{s_2^2}+m_2 \frac{y_2-y_3}{s_1^2}; 
\end{aligned}
\end{equation}
\begin{equation} \label{2_2}
\begin{aligned}
\pi \dv{y_1}{t}&=m_2 \frac{x_1-x_2}{s_3^2}-m_3 \frac{x_3-x_1}{s_2^2}; \\
\pi \dv{y_2}{t}&=m_3 \frac{x_2-x_3}{s_1^2}-m_1 \frac{x_1-x_2}{s_3^2}; \\
\pi \dv{y_3}{t}&=m_1 \frac{x_3-x_1}{s_2^2}-m_2 \frac{x_2-x_3}{s_1^2},
\end{aligned}
\end{equation}
and in polar coordinates
\begin{equation} \label{2_3}
\begin{aligned}
& \pi \dv{\varrho_1}{t}=-\frac{m_3 \varrho_2 \sin \left(\vartheta_1-\vartheta_3\right)}{s_3^2}+\frac{m_3 \varrho_3 \sin \left(\vartheta_3-\vartheta_1\right)}{s_2^2}; \\
& \pi \dv{\varrho_2}{t}=-\frac{m_3 \varrho_3 \sin \left(\vartheta_3-\vartheta_3\right)}{s_1^2}+\frac{m_1 \varrho_1 \sin \left(\vartheta_1-\vartheta_2\right)}{s_3^2}; \\
& \pi \dv{\varrho_3}{t}=-\frac{m_1 \varrho_1 \sin \left(\vartheta_3-\vartheta_1\right)}{s_2^2}+\frac{m_2 \varrho_3 \sin \left(\vartheta_2-\vartheta_3\right)}{s_1^2} ;
\end{aligned}
\end{equation}
\begin{equation} \label{2_4}
\begin{aligned}
& \pi \varrho_1 \dv{\vartheta_1}{t}=m_2 \frac{\varrho_1-\varrho_2 \cos \left(\vartheta_1-\vartheta_2\right)}{s_3^2}+m_3 \frac{\varrho_1-\varrho_3 \cos \left(\vartheta_3-\vartheta_1\right)}{s_2^2}; \\
& \pi \varrho_2 \dv{\vartheta_2}{t}=m_3 \frac{\varrho_2-\varrho_3 \cos \left(\vartheta_2-\vartheta_3\right)}{s_1^2}+m_1 \frac{\varrho_2-\varrho_1 \cos \left(\vartheta_1-\vartheta_2\right)}{s_2^2}; \\
& \pi \varrho_3 \dv{\vartheta_3}{t}=m_1 \frac{\varrho_3-\varrho_1 \cos \left(\vartheta_3-\vartheta_1\right)}{s_3^2}+m_2 \frac{\varrho_3-\varrho_2 \cos \left(\vartheta_2-\vartheta_3\right)}{s_3^2}.
\end{aligned}
\end{equation}
Here
\begin{equation} \label{2_5}
\begin{aligned}
& s_1^2=\left(x_2-x_3\right)^2+\left(y_2-y_3\right)^2=\varrho_2^2+\varrho_3^2-2 \varrho_2 \varrho_3 \cos \left(\vartheta_2-\vartheta_3\right); \\
& s_2^2=\left(x_3-x_1\right)^2+\left(y_3-y_1\right)^2=\varrho_3^2+\varrho_1^2-2 \varrho_3 \varrho_1 \cos \left(\vartheta_3-\vartheta_1\right); \\
& s_3^2=\left(x_1-x_2\right)^2+\left(y_1-y_2\right)^2=\varrho_1^2+\varrho_2^2-2 \varrho_1 \varrho_2 \cos \left(\vartheta_1-\vartheta_2\right).
\end{aligned}
\end{equation}

We first assume that $m_1 + m_2 + m_3$ is non-zero. The center of gravity of the three vortex threads can then be assumed to lie at the origin, and the first two equations of~\eqref{1_6}, which state that the center of gravity remains at rest, become
\begin{equation}\label{2_6}
  \begin{aligned}
    m_1 x_1 + m_2 x_2 + m_3 x_3 & = 0; \\
    m_1 y_1 + m_2 y_2 + m_3 y_3 & = 0,
  \end{aligned}
\end{equation}
or in polar coordinates
\begin{equation}\label{2_7}
  \begin{aligned}
    m_1 \varrho_1 \cos{\vartheta_1}+ m_2 \varrho_2 \cos{\vartheta_2} + m_3 \varrho_3 \cos{\vartheta_3}& = 0 \\
    m_1 \varrho_1 \sin{\vartheta_1}+ m_2 \varrho_2 \sin{\vartheta_2} + m_3 \varrho_3 \sin{\vartheta_3}& = 0
  \end{aligned}
\end{equation}
It is useful to write the third and fourth of the general integrals~\eqref{1_6} in the following form
\begin{align}
m_1 \varrho_1^2+m_2 \varrho_2^2+m_3 \varrho_3^2 & =C^{\prime}; \label{2_8}\\
\frac{1}{m_1} \log s_1+\frac{1}{m_2} \log s_2+\frac{1}{m_3} \log s_3 & =C. \label{2_9}
\end{align}
We multiply the first of equations~\eqref{2_7} by $\sin{\vartheta_1}$, $\sin{\vartheta_2}$, and $\sin{\vartheta_3}$, the second by $-\cos{\vartheta_1}$, $-\cos{\vartheta_2}$, and $-\cos{\vartheta_3}$ and add each time. In this way, three equations result, which can be written most simply as
\begin{equation} \label{2_10}
\frac{\sin \left(\vartheta_2-\vartheta_3\right)}{m_1 \varrho_1}=\frac{\sin \left(\vartheta_3-\vartheta_1\right)}{m_2 \varrho_2}=\frac{\sin \left(\vartheta_1-\vartheta_2\right)}{m_3 \varrho_3}.
\end{equation}

If one further brings the first term in the equations mentioned on the right side, then squares it and adds, then one obtains the first of the following equations, from which the other two result by cyclic permutation of the indices 1, 2, and 3, namely
\begin{equation} \label{2_11}
\begin{aligned}
& \cos \left(\vartheta_2-\vartheta_3\right)=\frac{m_1^2 \varrho_1^2-m_2^2 \varrho_2^2-m_3^2 \varrho_3^2}{2 m_2 m_3 \varrho_2 \varrho_3}; \\
& \cos \left(\vartheta_3-\vartheta_1\right)=\frac{-m_1^2 \varrho_1^2+m_2^2 \varrho_2^2-m_3^2 \varrho_3^2}{2 m_3 m_1 \varrho_3 \varrho_1}; \\
& \cos \left(\vartheta_1-\vartheta_2\right)=\frac{-m_1^2 \varrho_1^2-m_2^2 \varrho_2^2+m_3^2 \varrho_3^2}{2 m_1 m_2 \varrho_1 \varrho_2};
\end{aligned}
\end{equation}
We insert these expressions for $\cos{(\vartheta_2 - \vartheta_2)}$, $\cos{(\vartheta_3 - \vartheta_1)}$, and $\cos{(\vartheta_1 - \vartheta_2})$ into Eq.~\eqref{2_5}in Eq.~\eqref{2_5}, while using Eq.~\eqref{2_8}  and  arrive at the following formulas:
\begin{equation} \label{2_12}
\begin{aligned}
& m_2 m_3 s_1^2=\left(m_2+m_3\right) C^{\prime}-m_1\left(m_1+m_2+m_3\right) \varrho_1^2; \\
& m_3 m_1 s_2^2=\left(m_3+m_1\right) C^{\prime}-m_2\left(m_1+m_2+m_3\right) \varrho_2^2 ;\\
& m_1 m_2 s_3^2=\left(m_1+m_2\right) C^{\prime}-m_3\left(m_1+m_2+m_3\right) \varrho_3^2 .
\end{aligned}
\end{equation}
We introduce a new constant $C''$, which is related to $C'$ by the equation
\begin{equation} \label{2_13}
  m_1 m_2 m_3 C'' = (m_1 + m_2 + m_3) C'.
\end{equation}
It then follows from the previous equations that
\begin{equation} \label{2_14}
\frac{s_1^2}{m_1}+\frac{s_2^2}{m_2}+\frac{s_3^2}{m_3}=C'' .
\end{equation}

Equations~\eqref{2_9} and~\eqref{2_14} define have two integrals of the differential equations of our problem,  which depend only on the side lengths of the triangle formed by the three vortex threads. It is, therefore, reasonable to assume that three differential equations can be set up in a fairly simple manner, which only contain the time and the side lengths of the triangle $s_1$, $s_2$, and $s_3$, and from which the shape of the triangle can be determined at any moment. To create these equations, subtract the third equation from the second in Eqs.~\eqref{2_1} and~\eqref{2_2}, multiply the first of the resulting equations by $x_2-x_3$ the second by $y_2-y_3$ and add. This gives
\begin{equation*}
 \frac{\pi}2 \dv{\left(s_1^2\right)}{t}
  = m_1 \frac{s_2^2-s_3^2}{s_2^2 s_3^2}
  \left(
  y_1\left(x_2-x_3\right)+y_2\left(x_3-x_1\right)+y_3\left(x_1-x_2\right)
  \right).
\end{equation*}

We assume that the $xy$-axis system is chosen so that the positive $y$-axis is rotated by $90^\circ$ in the negative sense, i.e., clockwise, and lies where the $x$-axis originally was. The expression in brackets on the right-hand side of the above equation then represents twice the signed area of the triangle, depending on whether one must go around the triangle in a negative or positive sense to visit threads 1, 2, and 3 in sequence. This area can be expressed in a known way in terms of the side lengths. If we denote the above expression by $2 J$, then the first of equations~\eqref{2_15} results, and from this, the other two follow through the cyclic interchange of the indices 1, 2, and 3, namely
\begin{equation}
\begin{aligned} \label{2_15}
\dv{\left(s_1^2\right)}{t}&=\frac{m_1}{\pi} 4 J \frac{s_2^2-s_3^2}{s_2^2 s_3^2}; \\
\dv{\left(s_2^2\right)}{t}&=\frac{m_2}{\pi} 4 J \frac{s_3^2-s_1^2}{s_3^2 s_1^2} ;\\
\dv{\left(s_3^2\right)}{t}&=\frac{m_3}{\pi} 4 J \frac{s_1^2-s_2^2}{s_1^2 s_2^2} . \\
\end{aligned}
\end{equation}
The variable $J$ is determined by the equation
\begin{equation} \label{2_16}
16 J^2=2 s_2^2 s_3^2+2 s_3^2 s_1^2+2 s_1^2 s_2^2-s_1^4-s_2^4-s_3^4 .
\end{equation}
The two previously-determined integrals~\eqref{2_9} and~\eqref{2_14} result if equation~\eqref{2_15} is divided once by $m_1$, $m_2$, $m_3$, then by $m_1 s_1^2$, $m_2 s_2^2$, $m_3 s_3^2$, and the results added together.

Determining the triangle's shape at each instant requires only eliminations and squaring. The motion is completely determined if one still has an equation in which one or more coordinates and the time occur. Using~\eqref{2_11} and~\eqref{2_12}, the equations~\eqref{2_4} can be transformed in such a way that apart from one of the derivatives
$$
\dv{\vartheta_1}{t}, \dv{\vartheta_2}{t}, \dv{\vartheta_3}{t}
$$
only contain sides $s_1$, $s_2$, $s_3$. Namely, the following system of equations is obtained
\begin{equation} \label{2_17}
\begin{aligned}
& 2 \pi\left(m_1+m_2+m_3\right) \varrho_1^2 s_2^2 s_3^2 \dv{\vartheta_1}{t}&= 
 m_2 m_3\left\{\left(s_2^2-s_3^2\right)^2-s_1^2\left(s_2^2+s_3^2\right)\right\}+2\left(m_2+m_3\right)^2 s_2^2 s_3^2 \\
& 2 \pi\left(m_1+m_2+m_3\right) \varrho_2^2 s_3^2 s_1^2 \dv{\vartheta_2}{t}&= 
m_3 m_1\left\{\left(s_3^2-s_1^2\right)^2-s_2^2\left(s_3^2+s_1^2\right)\right\}+2\left(m_3+m_1\right)^2s_3^2 s_1^2 \\
& 2 \pi\left(m_1+m_2+m_3\right) \varrho_3^2 s_1^2 s_2^2 \dv{\vartheta_3}{t}
&=m_1 m_2\left\{\left(s_1^2-s_2^2\right)^2-s_3^2\left(s_1^2+s_2^2\right)\right\}+2\left(m_1+m_2\right)^2 s_1^2 s_2^2 
\end{aligned}
\end{equation}
in which the $\varrho$ variables are related to the $s$ variables by the equations~\eqref{2_12} and are only kept in these formulas for ease of writing. This group of differential equations only applies if the center of gravity is at the origin.

The present problem can now be solved in a general way as follows. From the equations~\eqref{2_8}, \eqref{2_9}, \eqref{2_11}, \eqref{2_12}, and \eqref{2_14}, the following nine variables
\begin{equation*}
\begin{array}{ccc}
s_1, & s_2, & s_3, \\
\varrho_1, & \varrho_2, & \varrho_3, \\
\cos \left(\vartheta_2-\vartheta_3\right), & \cos \left(\vartheta_3-\vartheta_1\right), & \cos \left(\vartheta_1-\vartheta_2\right)
\end{array}
\end{equation*}
can all be represented as functions of a single variable. Substituting the expressions thus obtained into any of equations~\eqref{2_3} or~\eqref{2_15}, squaring gives $t$ as a function of $\tau$ and inversely $\tau$ as a function of $t$. Using equations~\eqref{2_4} or~\eqref{2_17} one now also obtains the quantities $\vartheta_1$, $\vartheta_2$, $\vartheta_3$ as functions of time by squaring.

The above calculations can, however, be generalized, i.e., for arbitrary values of the constant $m$, and one must, therefore, confine oneself to integrating the differential equations of the problem only for a few very specially chosen sets of $m$ values. Equation~\eqref{2_9} is transcendental in general and algebraic only when the ratios of $m_1$, $m_2$, and $m_3$ are rational numbers.

The simplest assumptions that can be made about the $m$ are the following three
\begin{equation*}
  \begin{aligned}
  m_1 & = \phantom2m_2 & = - \phantom2 m_3,& \\
  m_1 & = \phantom2 m_2 & = \phantom{-2} m_3, &\qand \\
  m_1 & = 2 m_2 & = - 2 m_3, &\\
 \end{aligned} 
\end{equation*}
which we will revisit.

The previous calculation assumed that the center of gravity of the vortex threads coincides with the origin. This assumption is no longer valid if $m_1 + m_2 + m_3 = 0$ since the center of gravity is at infinity.

In this case, it is best to calculate in rectangular coordinates. One of the axes, e.g., the $x$-axis, can be chosen to coincide with the direction of motion of the center of gravity so that instead of equations~\eqref{2_6}, the following equations appear
$$
\begin{aligned}
  m_1 x_1 + m_2 x_2 + m_3 x_3 & = \text{const.}; \\
  m_1 y_1 + m_2 y_2 + m_3 y_3 & = 0.
\end{aligned}
$$
By appropriately choosing the coordinates, one can still cause the constant $C'$ in Eq.~\eqref{2_8} to disappear. However, the simplest conceivable assumption about the constant $m$, namely
$$
 -m_1 = 2 m_2 = 2 m_3,
$$
also leads to very complicated equations that defy detailed discussion. When we want solely to determine the shape of the triangle, we will use system~\eqref{2_15}, which is valid for all values of the quantities $m$ since they are independent of any coordinate system.

We now proceed to the treatment of the special cases mentioned above.

\section{The first case $m_1 = m_2 = -m_3$}

Equations~\eqref{2_5} imply
\begin{equation} \label{3_1}
  x_3 = x_1 + x_2, \quad y_3 = y_1 + y_2
\end{equation}
and demonstrate the theorem that the vortex threads and their center of gravity always form the corners of a parallelogram, with thread 3 and the center of gravity at opposite corners. Instead of the arbitrary constant $C'$ in Eq.~\eqref{2_8}, we introduce another constant $\lambda$ by writing $C' = 4 m_1 \lambda$. The above equation becomes 
\begin{equation} \label{3_2}
  \varrho_1^2 + \varrho_2^2 - \varrho_3^2 = 4 \lambda.
\end{equation}
It follows from Eq.~\eqref{2_12} that
\begin{equation} \label{3_3}
  s_1^2 = \varrho_1^2, \,
  s_2^2 = \varrho_2^2, \,
  s_3^2 = \varrho_3^2 + 8\lambda,
\end{equation}
and from Eq.~\eqref{2_9} that
\begin{equation*}
  \frac{\varrho_3^2 + 8\lambda}{\varrho_1^2 \varrho_2^2} = \text{const.}
\end{equation*}

Without loss of generality, one may assign a special value to this constant; such an assumption only determines a definition of the unit of length. We give the constant the value 1 so that one obtains 
\begin{equation} \label{3_4}
  \varrho_1^2 \varrho_2^2 -\varrho_3^2= 8 \lambda,
\end{equation}
From Eqs.~\eqref{3_2} and~\eqref{3_4} and the elimination of $\varrho_3$ results
\begin{equation} \label{3_5}
  (\varrho_1^2-1)(\varrho_2^2 -1) = 1+ 4\lambda,
\end{equation}
and it follows from this if we assume $1 + 4\lambda$ non-zero
\begin{equation} \label{3_6}
  \varrho_2^2 = \frac{\varrho_1^2 + 4\lambda}{\varrho_1^2-1}.
\end{equation}
If $1 + 4\lambda=0$ then from Eq.~\eqref{3_5}
$$
  (\varrho_1^2-1)(\varrho_2^2 -1) = 0
$$
In this equation, one can either make both factors or just one of them vanish.

If both factors are set equal to zero at the same time, the result is
\begin{equation} \label{3_7}
  \begin{aligned}
    \varrho_1 &= 1, & \varrho_2 & = 1, & \varrho_3 & = \sqrt{3}, \\
      s_1 &= 1, &  s_2 & = 1, &  s_3 & = 1;
  \end{aligned}
\end{equation}
the triangle formed by the three vortices remains equilateral and does not change in size. From the equations~\eqref{2_17}, it now follows that
\begin{equation}\label{3_8}
  \dv{\vartheta_1}{t} = \dv{\vartheta_2}{t} = \dv{\vartheta_3}{t} = \frac{m_1}{\pi},
\end{equation}
and the triangle rotates around the center of gravity with constant speed.

If one only wants to set one of the two factors equal to zero, then it is irrelevant whether one assumes $\varrho_1 = 1$ or $\varrho_2 = 1$ since swapping threads 1 and 2 is irrelevant; we want to assume $\varrho_2 = 1$ so that equation~\eqref{3_5} also remains valid for the case $\lambda = -\frac14$. The equation for $\varrho_3^2$ results from Eqs.~\eqref{3_4} and~\eqref{3_6}
\begin{equation} \label{3_9}
\varrho_3^2=\frac{\varrho_1^4-4 \lambda \varrho_1^2+8 \lambda}{\varrho_1^2-1} .
\end{equation}

Using these values, the last two equations of~\eqref{2_11} merge into the following 
\begin{equation} \label{3_10}
\begin{aligned}
 &\cos {\left(\vartheta_3-\vartheta_1\right)}
 &=\frac{\varrho_1^2-2 \lambda}{\varrho_1 \varrho_3}
 &=\frac{\varrho_1^2-2 \lambda}{\varrho_1} \sqrt{\frac{\varrho_1^2-1}{\varrho_1^4-4 \lambda \varrho_1^2+8 \lambda}}; \\
 &\cos {\left(\vartheta_1-\vartheta_2\right)}
 &=\frac{-2 \lambda}{\varrho_1 \varrho_2}
 &=\frac{-2 \lambda}{\varrho_1} \sqrt{\frac{\varrho_1^2-1}{\varrho_1^2+4 \lambda}}.
\end{aligned}
\end{equation}

Unless expressly stated otherwise, the quantities $\varrho$ should be positive. Therefore, the square roots in the previous equations are also to be taken as positive. For $\sin{ (\vartheta_3-\vartheta_1)}$, $\sin{ (\vartheta_1-\vartheta_2)}$ the following expressions result
\begin{equation} \label{3_11}
\begin{aligned}
& \sin \left(\vartheta_3-\vartheta_1\right)=\frac{1}{\varrho_1} \sqrt{\frac{\varrho_1^4-4\left(\lambda^2-\lambda\right) \varrho_1^2+4 \lambda^2}{\varrho_1^4-4 \lambda \varrho_1^2+8 \lambda}}; \\
& \sin \left(\vartheta_1-\vartheta_2\right)=-\frac{1}{\varrho_1} \sqrt{\frac{\varrho_1^4-4\left(\lambda^2-\lambda\right) \varrho_1^2+4 \lambda^2}{\varrho_1^2+4 \lambda}} .
\end{aligned}
\end{equation}
The sign of one of the two sines can be chosen arbitrarily; the sign of the other is then determined by Eq.~\eqref{2_10}.

A special value may also be attached to quantity $m_1$, which we want to assume is positive; it is through such an assumption that the unit of time is determined. We want to assume $m_1 = \pi$. Using the formulas developed so far, the first equations of systems~\eqref{2_3} and~\eqref{2_4} now merge into the following 
\begin{align}
 \dv{\varrho_1}{t}& =-\frac{\left(\varrho_1^2-1\right)^{3 / 2}\left(\varrho_1^4-4\left(\lambda^2-\lambda\right) \varrho_1^2+4 \lambda^2\right)^{1 / 2}}{\varrho_1^3\left(\varrho_1^2+4 \lambda\right)}, \label{3_12} \\
 \dv{\vartheta_1}{t}&=\frac{\left(\varrho_1^2-1\right)\left((1-2 \lambda) \varrho_1^2+2 \lambda\right)}{\varrho_1^4\left(\varrho_1^2+4 \lambda\right)}, \label{3_13}
\end{align}
and from these, it follows by the elimination of time
\begin{equation} \label{3_14}
\dv{\varrho_1}{\vartheta_1}=-\frac{\varrho_1\left(\varrho_1^2-1\right)^{1 / 2}\left(\varrho_1^4-4\left(\lambda^2-\lambda\right) \varrho_1^2+4 \lambda^2\right)^{1 / 2}}{(1-2 \lambda) \varrho_1^2+2 \lambda}.
\end{equation}
In these equations, one only needs to replace the subscript 1 with the subscript 2 to obtain the formulas for thread 2.

Before proceeding further with the above equations, let us determine the filaments' speeds. If $w$ denotes the speed of a moving point whose polar coordinates are $\varrho$ and $\vartheta$, then
\begin{equation*}
w^2=\left(\dv{\varrho}{t}\right)^2+\varrho^2\left(\dv{\vartheta}{t}\right)^2.
\end{equation*}

The equation for $w_1$ results from Eqs.~\eqref{3_12} and~\eqref{3_13}
\begin{equation} \label{3_15}
  w_1 = \frac{1}{\varrho_2^2},
\end{equation}
and from this, by swapping the indices 1 and 2
\begin{equation} \label{3_16}
  w_2 = \frac{1}{\varrho_1^2}.
\end{equation}
The last equations of systems~\eqref{2_3} and~\eqref{2_4} become the following with the use of Eqs.~\eqref{3_6} and~\eqref{3_9} etc.
\begin{align}
\dv{\varrho_3}{t}&=-\frac{\varrho_1^{4}-2 \varrho_1^2-4 \lambda}{\varrho_1^2\left(\varrho_1^2+4 \lambda\right)} \sqrt{\frac{\varrho_1^{4}-4\left(\lambda^2-\lambda\right) \varrho_1^2+4 \lambda^2}{\varrho_1^{4}-4 \lambda \varrho_1^2+8 \lambda}} \label{3_17} ;\\
\dv{\vartheta_3}{t}& =\frac{2\left(\varrho_1^2-1\right)\left((1-\lambda) \varrho_1^{4}+4 \lambda \varrho_1^2-4 \lambda^2\right)}{\varrho_1^2\left(\varrho_1^2+4 \lambda\right)\left(\varrho_1^{4}-4 \lambda \varrho_1^2+8 \lambda\right)}, \label{3_18}
\end{align}
yielding
\begin{equation} \label{3_19}
  w_3=\frac{\varrho_3}{\varrho_1 \varrho_2} .
\end{equation}

Equations~\eqref{3_12}, \eqref{3_13},~\eqref{3_17}, and ~\eqref{3_18} result in the maxima and minima of the quantities $\varrho$ and $\vartheta.$

From~\eqref{3_12}, and~\eqref{3_14}, one obtains $t$ and $\vartheta$ by elliptic integrals as functions of $\varrho$, specifically if we let
\begin{equation} \label{3_20}
  \varrho_1^2 = z
\end{equation}
 and set
\begin{align}
t & =\int \frac{z(z+4 \lambda)}{2(1-z)} \frac{\dd z}{\sqrt{(z-1)\left(z^2-4\left(\lambda^2-\lambda\right) z+4 \lambda^2\right)}} \label{3_21}\\
\vartheta_1 & =\int \frac{(2 \lambda-1) z-2 \lambda}{z} \frac{\dd z}{\sqrt{(z-1)\left(z^2-4\left(\lambda^2-\lambda\right) z+4 \lambda^2\right)}}. \label{3_22}
\end{align}

The equation
$$
z^2-4\left(\lambda^2-\lambda\right) z+4 \lambda^2=0
$$
has the roots
\begin{equation} \label{3_23}
  z_1=\left(\lambda-\sqrt{\lambda^2-2 \lambda}\right)^2, \quad z_2=\left(\lambda+\sqrt{\lambda^2-2 \lambda}\right)^2,
\end{equation}
which are identical for $\lambda = 0$ and for $\lambda = 2$. In this case, the integrals are then no longer elliptical but logarithmic. When $\lambda = - \frac{1}{4} $ then $z_1= 1$ and the same reduction occurs. These are the only three values of $\lambda$ that lead to logarithmic integrals; the corresponding motions will be treated in more detail.

In the reduction of the integrals~\eqref{3_21} and~\eqref{3_22}, one has to distinguish four cases according to the previous argument, namely
\label{lambda_cases}
\begin{enumerate}
  \item $-\infty<\lambda<-\frac{1}{4}$,
\item $-\frac{1}{4}<\lambda<0$
\item $0<\lambda<2$,
\item $2<\lambda<\infty$.
\end{enumerate}
In the four cases, the roots $z_1$ and $z_2$ lie within the following intervals
\begin{enumerate}
  \item $\infty>z_1>1, \quad 1>z_2>\frac{1}{4}$
  \item $1>z_1>0, \frac{1}{4}>z_2>0$
  \item $z_1$ and $z_2$ complex
  \item $4>z_1>1, \quad 4<z_2<\infty$.
\end{enumerate}

It suffices to carry out the reduction for one of these cases. We choose the second case. In descending order, the values for which the third-degree function under the square root sign vanishes are $1$, $z_1$, and $z_2$. The entire function is positive if $z$ lies between $z_1$ and $z_2$ or $z>1$. Only under the first assumption do we continue the calculation. Setting
\begin{equation} \label{3_24}
  z=z_2+\left(z_1-z_2\right) \sin^2{\psi},
\end{equation}
then
$$
\frac{\dd z}{\sqrt{(z-1)\left(z^2-4\left(\lambda^2-\lambda\right) z+4 \lambda^2\right)}}
=\frac{2}{\sqrt{1-z_2}} \frac{\dd\psi}{\sqrt{1-x^2 \sin^2{\psi}}},
$$
where
$$
x^2=\frac{z_1-z_2}{1-z_2}
$$
means a positive proper fraction. After carrying out some straightforward calculations, one obtains from Eqs.~\eqref{3_21} and~\eqref{3_22}
\begin{align} 
t&=\frac{-1}{\sqrt{1-z_2}}\left\{2(1+2 \lambda) F(x, \psi)-2 \frac{1+4 \lambda}{1-z_1} E(x, \psi) +\left(z_1-z_2\right) \frac{\sin{\psi} \cos{\psi}}{\sqrt{1-x^2 \sin^2{\psi}}}\right\} \label{3_25} \\
  \vartheta_1&=\frac{-1}{\sqrt{1-z_2}}\left\{(1-2 \lambda) F(x, \psi)+\frac{2 \lambda}{z_2} \Pi(x, \mu, \psi)\right\} \label{3_26}
\end{align}
where
$$
x^2=\frac{z_1-z_2}{1-z_2} \qand \mu=\frac{z_1-z_2}{z_2} .
$$

The functions $F(x, \psi), E(x, \psi), \Pi(x, \mu, \psi)$ are Legendre's incomplete elliptic integrals of the first, second, and third kind:
$$
\begin{gathered}
F(x, \psi)=\int_{0}^{\psi} \frac{\dd\psi}{\sqrt{1-x^2 \sin^2{\psi}}}, \quad 
E(x, \psi)=\int_{0}^{\psi} \sqrt{1-x^2 \sin^2{\psi}}\ \dd \psi , \\
\Pi(x, \mu, \psi)=\int_{0}^{\psi} \frac{\dd \psi}{\left(1+\mu \sin^2{\psi}\right) \sqrt{1-x^2 \sin^2{\psi}}} ;
\end{gathered}
$$
the constants of integration have been determined such that the quantities $t$, $\vartheta_1$, and $\psi$ vanish simultaneously. The integral of the third kind is always finite since the parameter $\mu$ is positive.

If one puts $\psi-\pi$ instead of $\psi$ in the above equations, the quantities $t$ and $\vartheta_1$ respectively increase by 
\begin{align}
 T&=\frac{4}{\sqrt{1-z_2}}\left\{(1+2 \lambda) K-\frac{1+4 \lambda}{1-z_1} E\right\} \label{3_27}\\
 \vartheta&=\frac2{\sqrt{1-z_2}}\left\{(1-2 \lambda) K+\frac{2 \lambda}{z_2} \Pi\right\} \label{3_28},
\end{align}
where $K$, $E$, and $\Pi$ mean the complete elliptic integrals of all three types; the quantities $\varrho$ and $s$ all remain unchanged. Therefore, The movement is periodic in that at time $t + T$, the threads are no longer in the same place as at time $t$ but in the same mutual position and state of motion.%
\footnote{Translator's footnote: This is called a relative periodic orbit in modern language.}
The trajectories of the threads are transcendental curves consisting of infinitely many congruent pieces. Namely, the trajectory of vortex 2 is the same curve as the trajectory of vortex 1, only rotated through the angle $\frac{1}{2}\vartheta$.

For $t=0$ we have
$$
\begin{gathered}
\varrho_1=\lambda+\sqrt{\lambda^2-2 \lambda}, \varrho_2=-\lambda+\sqrt{\lambda^2-2 \lambda}, \varrho_3=2 \sqrt{\lambda^2-2 \lambda}, \\
\vartheta_1=\vartheta_2=\vartheta_3=0,
\end{gathered}
$$
and for $t=\frac{1}{2} T$,
$$
\begin{gathered}
\varrho_1=-\lambda+\sqrt{\lambda^{2}-2 \lambda}, \varrho_2=\lambda+\sqrt{\lambda^{2}-2 \lambda}, \varrho_3=2 \sqrt{\lambda^2-2 \lambda}, \\
\vartheta_1=\vartheta_2=\vartheta_3=\frac{1}{2} \vartheta .
\end{gathered}
$$

\begin{figure}[htbp] 
  \centering
  \includegraphics[width=2in]{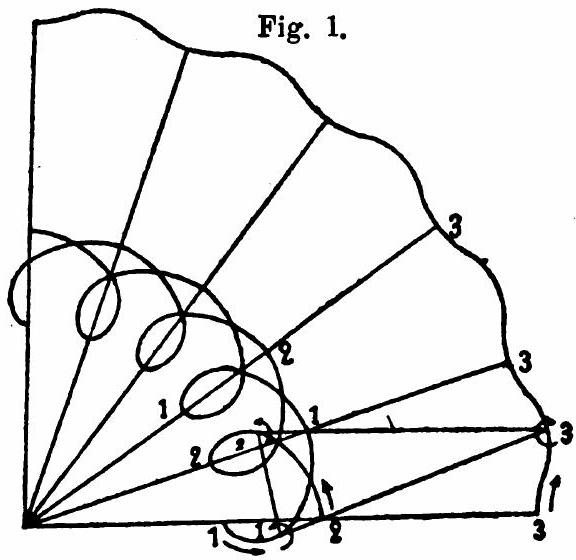} 
  \caption{Gröbli's Figure 1.}
  \label{fig:1}
\end{figure}

The motion for the interval $t=0$ to $t=\frac{1}{2} T$ is as follows. At the moment $t=0$, the three threads are in a straight line; $\varrho_1, \varrho_3$, as well as the velocity $w_1$ are minima, $\varrho_2$ as well as $w_2$ and $w_3$ are maxima. $\varrho_1$ and $v_1$ are now constantly growing, $\vartheta_1$ only decreases until $\varrho_1=\sqrt{\frac{2 \lambda}{2 \lambda-1}}$, then also to, at the time $t=\frac{1}{2} T$, $\varrho_1$ and $w_1$ have reached their greatest values, and $\vartheta_1$ has become $=\frac{1}{2} \vartheta$, while $\varrho_2$ and $w_2$ decrease continuously, and $\vartheta_2$ first increases, then decreases. At $t=\frac{1}{2} T$, $\varrho_ 2$ and $w_2$ are minima and $v_2=\frac{1}{2}\vartheta$. $\varrho_3$ finally increases $w_3$ decreases, at time $t=\frac{1}{4} T$ $\varrho_3$ is a maximum, $= (2-4 \lambda-2 \sqrt{1+4 \lambda})^{1 / 2}, \vartheta_3=\frac{1}{4} \vartheta, w_3$ a minimum. The triangle of the three vortices is isosceles at this moment. From this point on, $\varrho_3$ decreases and $w_3$ increases. At $t=\frac{1}{2} T$, $\varrho_3$ and $w_3$ have regained their original values. The three vortices are now again in the initial mutual position if one only interchanges 1 and 2, and with the same caveat, the motion for the interval $t=\frac{1}{2} T$ to $t=T$ is the same as for the one just described from $t=0$ to $t=\frac{1}{2} T$. Figure~\ref{fig:1}, which is based on the assumption $\lambda=-\frac{1}{12}$, should give an approximate picture of the course of the movement. The associated values of $T$ and $\vartheta$ are
for $t=0$
$$
T=0.1068, \quad \vartheta=0.6086 ;
$$ 
$$
\begin{aligned}
& \varrho_1&=\frac{1}3, \quad 
 \varrho_2&=\frac{1}{2}, \quad 
 \varrho_3&=\frac{5}{6} \\
&  s_1&=\frac{1}3, \quad 
   s_2&=\frac{1}{2}, \quad
   s_3&=\frac{1}{6} \\
&  w_1&=4, \quad
   w_2&=9, \quad
   w_3&=5 .
\end{aligned}
$$

Cases 1, 3, and 4 on page~\pageref{lambda_cases} can be treated similarly. However, the movement is no longer periodic; the time necessary for $z$ to pass from one extreme value to another diverges. Because either one of the limits of the integral is equal to 1 and for this value, the function under the integral sign in Eq.~\eqref{3_21} diverges to order $\frac32$, i.e., the integral itself of the order $\frac{1}{2}$, or $z=\infty$ is one of the limits; for this value of $z$, $\dv{t}{z}$ becomes zero of order $\frac{1}{2}$ and $t$ itself becomes infinite of this order. The angle ${\vartheta}_1$ only changes by a finite amount as $z$ varies through all possible values.

We proceed to the borderline cases mentioned above.

\section{The first borderline case $\lambda=0$}
\label{sec:4}
From~\eqref{2_14}, we have the result
\begin{equation} \label{4_29}
  s_3^2=s_1^2+s_2^2 ;
\end{equation}
the triangle of the three vortices is always right-angled. The integrations indicated in Eqs.~\eqref{3_21} and and~\eqref{3_22} can be carried out very easily. Making suitable choices of integration constants and again replacing $z$ by $\varrho_1^2$, one obtains
\begin{align}
t&=\frac{2-\varrho_1^2}{\sqrt{\varrho_1^2-1}} \label{4_30}\\
\vartheta_1&=-\operatorname{arctan} \sqrt{\varrho_1^2-1} . \label{4_31}
\end{align}

We want to assume the function arctangent in the first quadrant. Since $\dv{\varrho_1}{t}$ is continuously negative, $\dv{\vartheta_1}{t}$ must be positive, so $\varrho_1$ decreases and $\vartheta_1$ increases. From Eq.~\eqref{4_31} it follows that
\begin{equation} \label{4_32}
  x_1 = 1;
\end{equation}
thread 1 moves in a straight line parallel to the $y$ axis. From Eq.~\eqref{3_6} it follows that
$$\varrho_2=\frac{\varrho_1}{\sqrt{\varrho_1^2-1}}.
$$
By introducing rectangular coordinates into this equation and considering that, according to Eqs.~\eqref{3_10} and~\eqref{3_11},
$$
\vartheta_2-\vartheta_1=\frac{\pi}2,
$$
we find
\begin{equation} \label{4_33}
  x_2=1;
\end{equation}
thread 2 traverses the same straight line as thread 1. Finally, it follows from Eq.~\eqref{3_1} using the previous equation that
\begin{equation} \label{4_34}
  x_3=2;
\end{equation}
thread 3 also follows a straight trajectory line parallel to the $y$-axis. Since the motion is parallel to the $y$-axis, we introduce orthogonal coordinates, finding
$$
\varrho_1^2=1+y_1^2 ;
$$
inserting this expression into Eq.~\eqref{4_30} and taking into account that positive values of $t$ correspond to negative values of $y_1$, the result is
$$
t=\frac{y_1^2-1}{y_1}
$$
and from here
\begin{equation} \label{4_35}
  y_1=\frac{t-\sqrt{t^2+4}}2 .
\end{equation}
For $y_2$ and $y_3$ one obtains the formulas
\begin{align}
 y_2&=\frac{t+\sqrt{t^2+4}}2 \label{4_36}\\
 y_3&=t \label{4_37}.
\end{align}

Differentiating these equations with respect to $t$ yields these threads' velocities.

To summarize the obtained results, starting from the moment $t=0$, we can describe the movement as follows. Thread 1 moves in the straight line $x_1=1$. At time $t=0$ it is at the location $y_1=-1$, with speed  $\frac{1}{2}$. From this position, it moves parallel to the positive $y$-axis with steadily decreasing speed, getting closer and closer to the $x$-axis without ever reaching it. Thread 2 moves along the same straight line as 1 and continues from the initial position $y_2=1$ parallel to the positive $y$ axis. Its velocity increases continuously and converges to the limit 1. Finally, thread 3 moves along the straight line $x_3=2$ from the initial position $y_3=0$, with constant speed 1, in the same direction as the other two threads.

\section{The second borderline case $\lambda=-\frac{1}{4}$}
The following equations are obtained by integrating Eqs.~\eqref{3_21} and~\eqref{3_22}:
$$
\begin{aligned}
t & =-\frac{1}{2} \sqrt{4 z-1}+\frac{1}{\sqrt{3}} \log \left(\frac{\sqrt{4 z-1}+\sqrt{3}}{\sqrt{4 z-1}-\sqrt{3}}\right)+\text { const.}, \\
\vartheta_1 & =-\arctan\sqrt{4 z-1}+\frac{1}{\sqrt{3}} \log \left(\frac{\sqrt{4 z-1}+\sqrt{3}}{\sqrt{4 z-1}-\sqrt{3}}\right)+\text { const. }
\end{aligned}
$$

We must distinguish between the two  cases
$$
1<z<\infty, \qand \frac{1}{4}<z<1;
$$
reintroducing $\varrho_1^2$ in place of $z$  and suitably determining the constants of integration, one obtains the  equations
\begin{align}
 t=&-\frac{1}{2} \sqrt{4 \varrho_1^2-1}+\frac{1}{\sqrt{3}} \log \left(\frac{\sqrt{4 \varrho_1^2-1}+\sqrt{3}}{\sqrt{4 \varrho_1^2-1}-\sqrt{3}}\right), \label{5_38}
\\
 \vartheta_1=&-\arctan \sqrt{4 \varrho_1^2-1}+\frac{1}{\sqrt{3}} \log \left(\frac{\sqrt{4 \varrho_1^2-1}+\sqrt{3}}{\sqrt{4 \varrho_1^2-1}-\sqrt{3}}\right)  \qfor 1<\varrho_1<\infty, \label{5_39}
\end{align}
and
\begin{align}
 t & = -\frac{1}{2} \sqrt{4 \varrho_1^2 -1} + \frac{1}{\sqrt{3}} 
    \log{\left( \frac{\sqrt{3}+\sqrt{4\varrho_1^2-1}}{\sqrt{3}+\-\sqrt{4\varrho_1^2-1}}\right)}, \label{5_40} \\
 \vartheta_1 & = -\arctan{ \sqrt{4 \varrho_1^2 -1}} + \frac{1}{\sqrt{3}} 
    \log{\left( \frac{\sqrt{3}+\sqrt{4\varrho_1^2-1}}{\sqrt{3}+\-\sqrt{4\varrho_1^2-1}}\right)} \qfor \frac{1}{2}<\varrho_1<1. \label{5_41} 
\end{align}

From Eqs.~\eqref{3_3},~\eqref{3_6},~\eqref{3_9}, and~\eqref{3_15} it results that
\begin{equation} \label{5_42}
  \varrho_2=1, \quad s_2=1, \quad w_1=1, \quad s_1=s_3 .
\end{equation}
Accordingly, thread 2 moves in a circle whose center is the center of gravity of the three vortex threads; thread 1 traverses its path at a constant speed; the three vortices form the corners of an isosceles triangle whose base has constant length. For $\vartheta_2$ and $\vartheta_3$ the following equations result from~\eqref{3_9},~\eqref{3_10}, ~\eqref{3_11},~\eqref{5_39} and~\eqref{5_41}:
\begin{align}
  \vartheta_2& =\frac{1}{\sqrt{3}} \log \left( \pm \frac{\sqrt{4 \varrho_1^2-1}+\sqrt{3}}{\sqrt{4 \varrho_1^2-1}-\sqrt{3}}\right); \label{5_43} \\
  \vartheta_3 &=-\arctan{ \frac{1}3 \sqrt{4 \varrho_3^2-9}}+\frac{1}{\sqrt{3}} \log \left( \pm \frac{\sqrt{4 \varrho_3^2-9}+\sqrt{3}}{\sqrt{4 \varrho_3^2-9}-\sqrt{3}}\right), \label{5_44}
\end{align}
in which the positive or negative sign is to be taken under the sign of the logarithm, depending on whether $\varrho_1 \leq 1$. To discuss these equations further, we must distinguish between the cases $\varrho_1>1$ and $\varrho_1<1$.

\paragraph{1)} $\mathbf{1<\varrho_1<\infty}$\ \ Here $\varrho_3$ lies between $\sqrt{3}$ and $\infty$, $\varrho_1$ and $\varrho_3$ decrease continuously, whereas $\vartheta_1$, $\vartheta_2$, $\vartheta_3$, $w_2$, and $w_3$ increase; the threads move around the center of gravity in a positive sense. Equation~\eqref{5_39} can be written
$$
\vartheta_1+\frac{\pi}{2}=\arctan\frac{1}{\sqrt{4 \varrho_1^2-1}}+\frac{1}{\sqrt{3}} \log \left(\frac{\sqrt{4 \varrho_1^2-1}+\sqrt{3}}{\sqrt{4 \varrho_1^2-1}-\sqrt{3}}\right) .
$$
If one expands the right-hand side to powers of $\frac{1}{\varrho_1}$, assuming that $\varrho_1$ is very large and only keeps the terms of the first order, the result is
$$
\vartheta_1+\frac{\pi}{2}=\frac{3}{2 \varrho_1}
$$
or
$$
x_1=\frac{3}{2} .
$$
The straight line $x_1=\frac{3}{2}$ is an asymptote of the path of thread 1. It also follows from Eq.~\eqref{5_44} that the straight line
$$
x_3=\frac{5}{2}
$$
is an asymptote of the curve traversed by thread 3.

The trajectory of thread 1 is a spiral, which has the line $x_1=\frac{3}{2}$ and the circle of radius 1 as asymptotes; the trajectory of thread 3 is a spiral, for which the line $x_3=\frac{5}{2}$ and the circle of radius $\sqrt{3}$ are asymptotes. Both spirals rapidly approach their asymptotic circles.

For $t=-\infty$ we get
$$
\varrho_1=\infty, \vartheta_1=-\frac{\pi}{2}, \vartheta_2=0, \quad \varrho_3=\infty, \quad \vartheta_3=-\frac{\pi}{2},
$$
and for $t=\infty$
$$\varrho_1=1, \quad \vartheta_1=-\frac{\pi}{3}+\vartheta_2, \quad \vartheta_2=\infty, \quad \varrho_3=\sqrt{3}, \quad \vartheta_3=-\frac{\pi}{6}+\vartheta_2.
$$

Let us assume that threads 1 and 3 are very far from the center of gravity at the beginning of the movement. Then thread 2 is very close to the point $x_2=1, y_2=0$, the velocity $w_2$ is very small, and $w_3$ is almost equal to 1. Both velocities now increase rapidly and converge towards the values 1 and $\sqrt{3}$ without ever reaching them. After thread 2 has traversed its circle once, threads 1 and 2 move approximately along the circle of radius 1 with speed 1, while 3 moves approximately along the circle of radius $\sqrt{3}$ with speed $\sqrt{3}$. The triangle of the three threads is equilateral and rotates about the center of gravity with constant velocity. 

\paragraph{2)} $\mathbf{\frac{1}{2}<\varrho_1<1}$\ \ Now $\varrho_3$ lies between $\frac{3}{2}$ and $\sqrt{3}$. Thread 1 moves along a spiral whose minimal radius is $\frac{1}{2}$, for which the circle of radius 1 is an asymptotic circle and the $x$-axis is the symmetry axis. Thread 2 moves on the circle of radius 1; thread 3 describes a spiral, which has the $x$-axis to the axis of symmetry and the circle of radius $\sqrt{3}$ to the asymptotic circle. The smallest radius vector is $\frac{3}{2}$. Both spirals very quickly connect to their asymptotic circles. At the moment $t=0$, all three threads lie on the $x$-axis, at the distances $\frac{1}{2},$ $1,$ and $\frac{3}{2}$ from the centroid.

\begin{figure}[htbp] 
  \centering
  \includegraphics[width=3in]{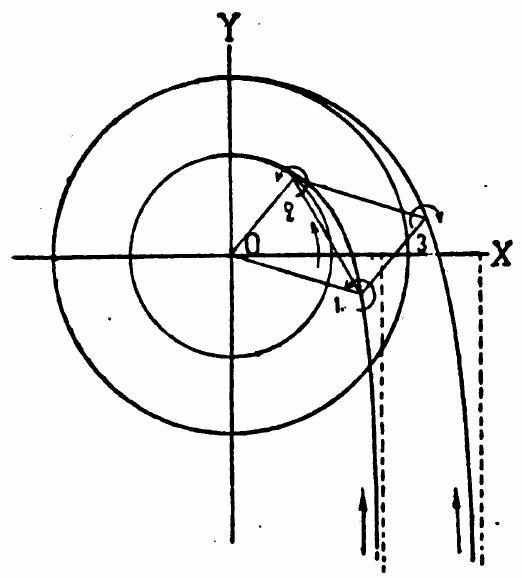} 
  \caption{Gröbli's Figure 2.}
  \label{fig:2}
\end{figure}

The speeds are $1$, $4$, and $3$ respectively. The angle $\vartheta_1$ first decreases, reaches the minimum for $\varrho_1=\frac{1}{3} \sqrt{3}$ and then increases steadily. $\varrho_3$, $\vartheta_1,$ and $\vartheta_3$ also increase. Thread 1 moves with the speed 1, while threads 2 and 3 trace their orbits with decreasing speed and converge towards the limits $1$ and $\sqrt{3}$. When thread 2 has traversed its circle once, which has happened after a certain finite time, the motion occurs nearly as if the triangle of the three threads were equilateral and rotating about the center of gravity at speed 1. o 

\begin{figure}[htbp] 
  \centering
  \includegraphics[width=3in]{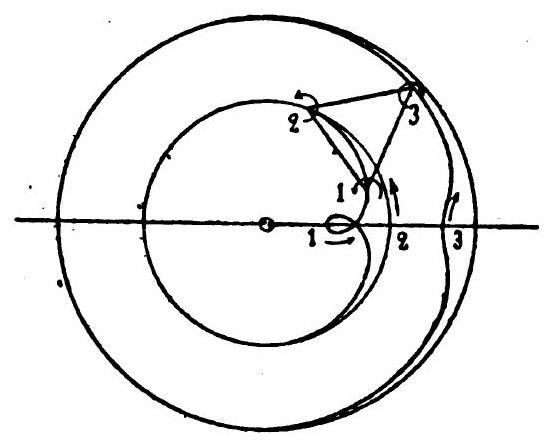} 
  \caption{Gröbli's Figure 3.}
  \label{fig:3}
\end{figure}

Figures~\ref{fig:2} and~\ref{fig:3} are intended to give an approximate idea of the course of the movement.

The assumption $\lambda=2$ leads to very similar motions. We omit the details.

\section{The Second case: $m_1=m_2=m_3$}

Equations~\eqref{2_8} and~\eqref{2_9} can be written
\begin{align}
\varrho_1^2+\varrho_2^2+\varrho_3^2 & =1, \label{6_1} \\
s_1 s_2 s_3 & =\lambda \label{6_2}.
\end{align}

The unit of length is suitably chosen so that $\lambda$ means a positive constant; otherwise, it may be arbitrary.

System~\eqref{2_12}, which conveys the relationship between the variables $\varrho$ and $s$, becomes
\begin{equation} \label{6_3}
  s_1^2=2-3 \varrho_1^2, \quad
  s_2^2=2-3 \varrho_2^2, \quad
  s_3^2=2-3 \varrho_3^2 .
\end{equation}

If you add these three equations, you get, using Eq.~\eqref{6_1},
\begin{equation} \label{6_4}
  s_1^2+s_2^2+s_3^2=3.
\end{equation}

With the help of Eqs.~\eqref{6_2} and~\eqref{6_4}, one can already get a general view of the motion, as far as only the shape of the triangle is considered, and a classification of the various possible cases can be set up. If we think of $s_1$, $s_2$, and $s_3$ as right-angled coordinates of a point in space, Eq.~\eqref{6_4} represents a sphere of radius $\sqrt{3}$ and Eq.~\eqref{6_2} a third-order surface which has the coordinate planes as asymptotic planes and is intersected by planes parallel to them in equilateral hyperbolas. Since the quantities $s_1$, $s_2$, and $s_3$ are positive, we can limit ourselves to the first octant. The levels
$$
s_2 = s_3, \; s_3= s_1, \qand s_1 = s_2
$$
are planes of symmetry for the sphere and the surface of the third order, and therefore also for the intersection curve of both, an oval-like figure. Each point of this curve corresponds to a specific shape of the triangle of the three threads. The oval has a highest and a lowest point, corresponding to the two extreme values between which $s_3$ must lie. Of course, $s_1$ and $s_2$  lie within the same boundaries. Since the plane $s_1=s_2$ is a plane of symmetry, then $s_3$ certainly obtains its maximum and minimum when $s_1=s_2$. There are no other maxima and minima. If we set $s_1=s_2$, we get an equation of the third degree for $s_3$
\begin{equation} \label{6_5}
  s_3^{3}-3 s_3+2 \lambda=0 .
\end{equation}
One root of this equation is always real but cannot be used because it is negative. The other two roots give the maximum and minimum values of $s_3$, respectively. They can be real and different, real and equal, or imaginary, depending on whether the two surfaces intersect, touch, or don't intersect. The contact occurs for $\lambda=1$ where the intersection curve is reduced to the point
\begin{equation} \label{6_6}
  s_1 = s_2 = s_3 = 1
\end{equation}
and the three threads form an equilateral triangle of constant dimension. From Eq.~\eqref{6_3} it follows that
$$
\varrho_1^2 = \varrho_2^2 = \varrho_3^2 = \frac{1}{3}.
$$

If we denote the common value of the quantities $m_1$, $m_2$, and $m_3$ by $m$, then Eq.~\eqref{2_17} results in 
\begin{equation} \label{6_7}
  \dv{\vartheta_1}{t}=\dv{\vartheta_2}{t}=\dv{\vartheta_3}{t}=\frac{3 m}{\pi} .
\end{equation}
Therefore, the triangle of three threads revolves at a constant speed around its center.

We now summarize what has been said so far. The sides $s_1$, $s_2$, and $s_3$ of the triangle of the three vortex threads have to satisfy Eqs.~\eqref{6_2} and~\eqref{6_4}, in the former $\lambda$ means a positive, non-zero constant lying between 0 and 1. Owing to these conditions, each side can only vary between two definite finite limits, which are the same for all three sides and are defined as the positive roots of the equation
$$
 s_3^{3}-3 s_3+2 \lambda=0 .
$$
When one of the sides has reached its extreme value, the triangle is isosceles.

Now, the following has to be considered. For the triangle to be real, the sum of two sides must be greater than the third. This condition is certainly met if one of the sides, e.g., $s_3$, is a minimum because the smaller root of the above equation is less than 1, the associated values of $s_1$ and $s_2$ are greater than 1 after Eq.~\eqref{6_4}, so $s_1+s_2>s_3$. On the other hand, if $s_3$ is a maximum, then it is greater than 1, so $s_1$ and $s_2$ are smaller than 1, and it depends entirely on the value of $\lambda$, whether $s_1+s_2 \gtreqless s_3$. The limiting case $s_1+s_2=s_3$ occurs for $s_3=\sqrt{2}$; the third degree equation gives $\sqrt{\frac{1}{2}}$ as the associated value of $\lambda$. In one border position, the three threads are in a straight line, one exactly midway between the other two. The three equations
$$
s_2+s_3=s_1, s_3+s_1=s_2, \text{ and } s_1+s_2=s_3
$$
represent the planes formed by two of the bisectors of the positive coordinate axes. These planes intersect the sphere in an equilateral spherical triangle, and the three cases $\lambda^2 \gtreqless \frac{1}{2}$ are distinguished in that the oval lies entirely inside the triangle or touches it or intersects it. In the last case, only an isosceles triangle shape is possible; another possible position is that the three threads are in a straight line, but not one midway between the other two.

We now determine the motion and will use the differential equations~\eqref{2_15} and~\eqref{2_17} for this purpose. Equations~\eqref{6_2} and~\eqref{6_4} initially result in
\begin{equation} \label{6_8}
  \begin{aligned}
& s_2+s_3=\sqrt{\frac{-s_1^{3}+3 s_1+2 \lambda}{s_1}}; \\
& s_2-s_3=\sqrt{\frac{-s_1^{3}+3 s_1-2 \lambda}{s_1}}; \\
& s_2^2-s_3^2=\frac{\sqrt{s_1^{6}-6 s_1^{4}+9 s_1^2-4 \lambda^2}}{s_1} .
\end{aligned}
\end{equation}
Using these formulas, one obtains from Eq.~\eqref{2_16}
\begin{equation} \label{6_9}
  4 J=\frac{\sqrt{-4 s_1^{6}+12 s_1^{4}-9 s_1^2+4 \lambda^2}}{s_1},
\end{equation}
and now from the first equations of systems~\eqref{2_15} and~\eqref{2_17} with a suitable choice of the time unit
\begin{align}
 \dv{\left(s_1^2\right)}{t} 
  & = \frac{1}{\lambda^2} \sqrt{\left(s_1^{6}-6 s_1^{4}+9 s_1^2-4 \lambda^2\right)\left(-4 s_1^{6}+12 s_1^{4}-9 s_1^2+4 \lambda^2\right)}; \label{6_10} \\
 \dv{\vartheta_1}{t} &=
 \frac{2 s_1^{6}-9 s_1^{4}+9 s_1^2+4 \lambda^2}{2 \lambda^2\left(2-s_1^2\right)} . \label{6_11}
\end{align}
 By eliminating $t$ from Eqs.~\eqref{6_10} and~\eqref{6_11} it follows that
\begin{equation} \label{6_12}
 \dv{\left(s_1^2\right)}{\vartheta_1}=\frac{2\left(2-s_1^2\right) \sqrt{\left(s_1^{6}-6 s_1^{4}+9 s_1^2-4 \lambda^2\right)\left(-4 s_1^{6}+12 s_1^{4}-9 s_1^2+4 \lambda^2\right)}}{2 s_1^{6}-9 s_1^{4}+9 s_1^2+4 \lambda^2} .
\end{equation} 

From~\eqref{6_10} and~\eqref{6_12}, we may obtain formulas for $t$ and $\vartheta_1$ as hyperelliptic integrals as functions of $s_1^2$. In particular, setting
\begin{equation} \label{6_13}
  s_1^2 = z
\end{equation}
yields
\begin{align}
t & =\int \frac{\lambda^2 d z}{\sqrt{\left(z^{3}-6 z^2+9 z-4 \lambda^2\right)\left(-4 z^{3}+12 z^2-9 z+4 \lambda^2\right)}} ;\label{6_14}\\
\vartheta_1 &=\int \frac{\left(2 z^{3}-9 z^2+9 z+4 \lambda^2\right) d z}{2(2-z) \sqrt{\left(z^{3}-6 z^2+9 z-4 \lambda^2\right)\left(-4 z^{3}+12 z^2-9 z+4 \lambda^2\right)}} \label{6_15}.
\end{align}
In the second of these equations, one only needs to express $z$ in terms of $\varrho_1^2$ with the help of Eq.~\eqref{6_3} to obtain the equation of the trajectory of thread 1.

The above equations are also valid for threads 2 and 3 if e the characters $s_1, \vartheta_1$ are replaced with $s_2, \vartheta_2$, or $s_3, \vartheta_3$.

The following equations are obtained for the speeds at which the threads move
\begin{equation}\label{6_16}
  w_1=\frac{3 \varrho_1 s_1}{\lambda}, \quad w_2=\frac{3 \varrho_2 s_2}{\lambda}, \quad w_3=\frac{3 \varrho_3 s_3}{\lambda} .
\end{equation}

In Equations~\eqref{6_14} and~\eqref{6_15}, the function of the sixth degree under the root sign is
\begin{equation} \label{6_17}
  f(z)=
    \left(z^{3}-6 z^2+9 z-4 \lambda^2\right)
    \left(-4 z^{3}+12 z^2-9 z+4 \lambda^2\right) .
\end{equation}
The equation $f(z)=0$ has double roots only if $\lambda^2$ has one of the values $0, \frac{1}{2}$, or 1. We can disregard the case $\lambda=0$ since it leads back to two vortex threads. We have already dealt with the case $\lambda=1$. Finally, we will consider the case $\lambda^2=\frac{1}{2}$ later. If $\lambda$ is different from one of the numbers $0, \sqrt{\frac{1}{2}}, 1$, then none of the six linear factors into which $f(z)$ can be decomposed are equal to each other, and the integrals in~\eqref{6_14} and~\eqref{6_15} are hyperelliptic. The integral in~\eqref{6_14}, therefore, always remains finite. In Eq.~\eqref{6_15}, there is a rational function of $z$ next to the root of the function of the sixth degree, which becomes infinitely large for $z=2$. But since $z$ can never equal 2, this integral is always finite. The movement is, therefore, periodic in such a way that after a certain finite time has elapsed, the threads are no longer in their original place but in the same mutual position and state of motion. The angle $\vartheta_1$ keeps growing. We must now distinguish between the cases $\lambda^2 \gtrless \frac{1}{2}$.

\paragraph{Case 1: $\lambda^2>\frac{1}{2}$.}
The equation
$$
z^{3}-6 z^2+9 z-4 \lambda^2=0
$$
has three real positive roots $z_1, z_3, z_3$ within the following intervals
\begin{equation} \label{6_18}
  2-\sqrt{3}<z_1<1, \quad 1<z_2<2, \qand 2+\sqrt{3}<z_3<4.
\end{equation}
The equation
$$
-4 z^{3}+12 z^2-9 z+4 \lambda^2=0
$$
has only one real root between 2 and 3. The values $z_1$ and $z_2$ correspond to the minimum and maximum of $s_1$, so $z$ must be between $z_1$ and $z_2$.

The time that is necessary for the three threads to return from a certain position to the same mutual position and the same state of motion, is according to Eq.~\eqref{6_14},
\begin{equation} \label{6_19}
  T=2 \lambda^2 \int_{z_1}^{z_2} \frac{\dd z}{\sqrt{f(z)}}.
\end{equation}
The angles $\vartheta_1$, $\vartheta_2$, and $\vartheta_3$ have increased during this time by
\begin{equation} \label{6_20}
  \vartheta=\int_{z_1}^{z_2} \frac{\left(2 z^{3}-9 z^2+9 z+4 \lambda^2\right) d z}{(2-z) \sqrt{f(z)}}.
\end{equation}
Let us imagine the movement beginning at the moment when the triangle formed by the three threads is isosceles with thread 1 at the apex and $s_1$ a minimum. The triangle now rotates around the center of gravity in a positive sense, changing its shape simultaneously. Side lengths $s_1$ and $s_3$ increase, and$s_2$ decreases. At the time $t=\frac{1}{2} T$, the triangle is isosceles again, and now thread 3 sits at the top, with $s_3$  a maximum, so the triangle is less acute than before. The triangle continues to rotate, $s_1$ increases, $s_2$ and $s_3$ decrease, and at time $t=\frac{1}{3} T$, the triangle again has its initial isosceles shape, only now thread 2 is at the top.  According to Eq.~\eqref{6_3}, a maximum or minimum of one of the sides always corresponds to a minimum or maximum of the distance of the thread opposite the relevant side from the center of gravity. The threads describe certain curves consisting of an infinite number of congruent pieces. If one rotates the path of one of the threads by the angle $\frac{2 \pi-\vartheta }{3}$ forwards and backward, one obtains the paths of the other two threads. For $\lambda=1$, $\vartheta=0$, and for $\lambda=\sqrt{\frac{1}{2}}$, $\vartheta=\infty$. There are therefore innumerable values of $\lambda$ for which $\frac{2 \pi-\vartheta}{3}$ is a multiple of $2 \pi$. In such a case, the three threads move forward on the same curve, alternately approaching and receding from one another.

The speed $w_1$ viewed as a function of time is a minimum for $t=0$, then increases and reaches a maximum of $\frac{ \sqrt{3}}{\lambda}$ after a certain time; at this moment $s_1=1$; then decreases, is a minimum for $t=\frac{1}{2} T$, increases again to $\frac{\sqrt{3}}{\lambda}$, and finally decreases to reach the initial value at $t=T$.

\begin{figure}[htbp] 
  \centering
  \includegraphics[width=0.5\textwidth]{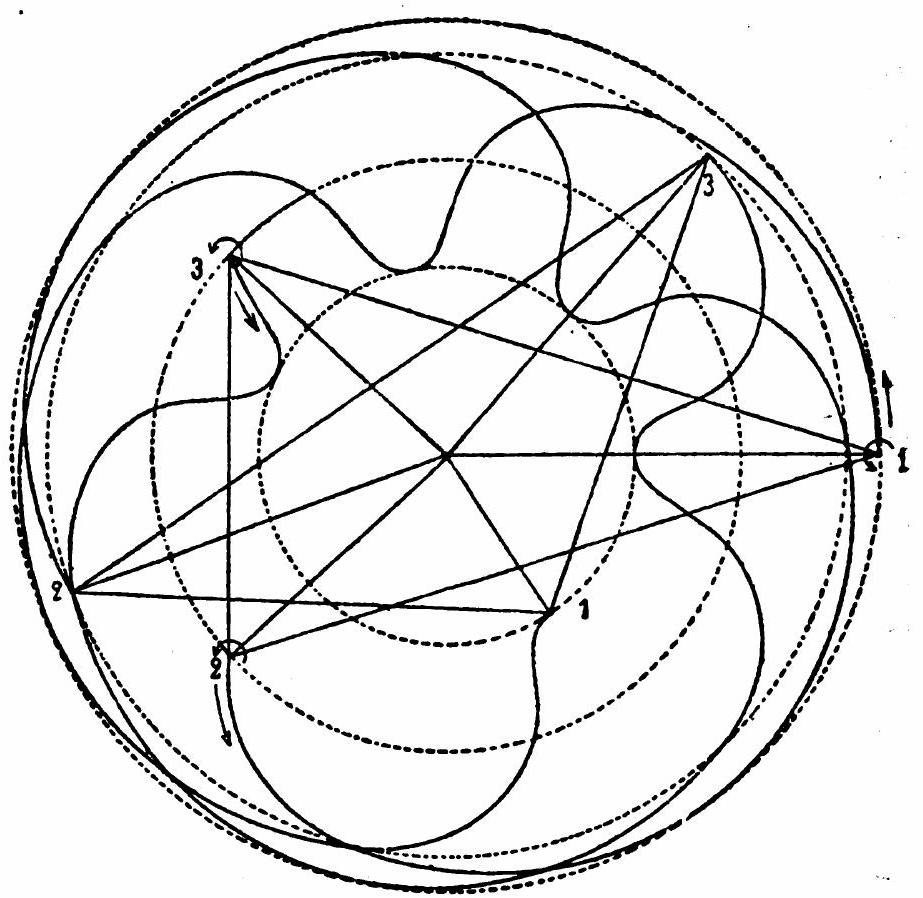} 
  \caption{Gröbli's Figure 4}
  \label{fig:4}
\end{figure}

Figure~\ref{fig:4} corresponds to the assumption $\lambda^2=\frac{243}{343}$. For this value of $\lambda^2$ one obtains
$$
\begin{aligned}
T& =2 \lambda^2 \int_{\frac{6}{14}}^{\frac{24}{14}} \frac{\dd z}{\sqrt{f(z)}}=6 \lambda^2 \int_{\frac{6}{14}}^{\frac{9}{14}} \frac{\dd z}{\sqrt{f(z)}} \approx 2.1078; \\
\vartheta&=\int_{\frac{6}{14}}^{\frac{24}{14}} \frac{\left(2 z^{3}-9 z^2+9 z+4 \lambda^2\right) \dd z}{(2-z) \sqrt{f(z)}} ;\\
&=3 \int_{\frac{6}{14}}^{\frac{9}{14}} \frac{\left(2 z^{3}-9 z^2+9 z+4 \lambda^2\right) \dd z}{(2-z) \sqrt{f(z)}}+3 \arccos \frac{17}{19}-\pi \approx 3.5355,
\end{aligned}
$$
and straightforwardly calculates the data in Table~\ref{table:1}, according to which the drawing is made.

\begin{table}[h]
  \centering
\begin{tabular}{|c||c||c|c|c||c|c|c|}
\hline $6 \frac{t}{T}$ & $6 \vartheta_1$ & $14 s_1^2$ & $14 s_2^2$ & $14 s_2^2$ & $42 \varrho_1^2$ & $42 \varrho_2^2$ & $42 \varrho_3^2$ \\
\hline \hline 0 & 0 & 6 & 18 & 18 & 22 & 10 & 10 \\
\hline 1 & $\vartheta+\pi-3 \arccos \frac{17}{19}$ & 9 & 9 & 24 & 19 & 19 & 9 \\
\hline 2 & $2 \vartheta+2 \pi-3 \arccos \frac{1}{10}$ & 18 & 6 & 18 & 10 & 22 & 10 \\
\hline 3 & $3 \vartheta$ & 24 & 9 & 9 & 4 & 19 & 19 \\
\hline 4 & $4 \vartheta-2 \pi+3 \arccos \frac{1}{10}$ & 18 & 18 & 6 & 10 & 10 & 22 \\
\hline 5 & $5 \vartheta-\pi+3 \arccos \frac{17}{19}$ & 9 & 24 & 9 & 19 & 4 & 19 \\
\hline 6 & $6 \vartheta$ & 6 & 18 & 18 & 22 & 10 & 10 \\
\hline
\end{tabular}
  \caption{Numerical values used in the creation of Figure~\ref{fig:4}.}
  \label{table:1}
\end{table}

\paragraph{Case 2: $\lambda^2<\frac{1}{2}$}

The equation
$$
z^{3}-6 z^2+9 z-4 \lambda^2=0
$$
has three real roots within the following bounds
\begin{equation} \label{6_21}
  0<z_1<2-\sqrt{3}, \quad 2<z_2<3, \quad 3<z_3<2+\sqrt{3} . 
\end{equation}
The root $z_1$ corresponds to the minimum of $s_1$, at which point the values of $z_2$ and $z_3$ are incompatible with a real triangle. The equation
$$
-4 z^{3}+12 z^2-9 z+4 \lambda^2=0
$$
also has three real positive roots within the bounds
\begin{equation} \label{6_22}
 0<z^{\prime}<\frac{1}{2}, \;
 \frac{1}{2}<z^{\prime \prime}<\frac{3}{2}, \; 
 \frac{3}{2}<z^{\prime \prime \prime}<2,  
\end{equation}
all three of which are compatible with a real triangle. The six roots of the equation $f(z)=0$ are, in order of magnitude,
$$
z_1<z^{\prime}<z^{\prime \prime}<z^{\prime \prime \prime}<z_2<z_3,
$$
and we may factor
$$
f(z)=-4\left(z-z_1\right)\left(z-z^{\prime}\right)\left(z-z^{\prime \prime}\right)\left(z-z^{\prime \prime \prime}\right)\left(z-z_2\right)\left(z-z_3\right),
$$
which is positive if $z$ is between $z_1$ and $z^{\prime}$, or between $z^{\prime \prime}$ and $z^{\prime \prime \prime}$, or finally lies between $z_2$ and $z_3$. $z=z^{\prime \prime \prime}$ is the largest value that $z$ can take on, so it must either be that $z_1 \leq z \leq z^{\prime}$, or $z^{\prime \prime} \leq z \leq z^{\prime \prime \prime}$. The difference between the two cases is that thread 1 has been swapped with one of threads 2 and 3. Let us assume that $z$ is between $z_1$ and $z^{\prime}$. The time that must elapse for the three threads to return to their initial mutual position and the same state of motion is
\begin{equation} \label{6_23}
  T=4 \lambda^2 \int_{z_1}^{z^{\prime}} \frac{ \dd z}{\sqrt{f(z)}},
\end{equation}
and during this time, the angles $\vartheta$ have increased by 
\begin{equation} \label{6_24}
  \vartheta=2 \int_{z_1}^{z^{\prime}} \frac{\left(2 z^{3}-9 z^2+9 z+4 \lambda^2\right) \dd z}{(2-z) \sqrt{f(z)}}. 
\end{equation}
The squared side length $s_1^2$ is bounded between the values $z_1$ and $z^{\prime}$, squared lengths $s_2^2$ and $s_3^2$ are within the boundaries $z^{\prime \prime}$ and $z^{\prime \prime \prime}$. We want to start when the triangle is isosceles, with thread 1 at the apex and $s_1$ a minimum. Then the lengths $s_1$ and $s_3$ increase, and $s_2$ decreases, until at time $t=\frac{1}{4} T$, the three threads are in a straight line, with thread 3 between thread 1 and thread 2, closer to thread 2 than to thread 1. The triangle continues to evolve, returning to its original shape at time $t=\frac{1}{2} T$. At time $t=\frac{3}{4} T$, the three threads are again in a straight line and indeed in the previous mutual position, only with thread 2 exchanged for thread 3, and so on. The curves traversed by threads 2 and 3 can be made to coincide by rotating about the center of gravity.

It remains to treat the case $\lambda^2=\frac{1}{2}$. The integrals in Eqs.~\eqref{6_14} and~\eqref{6_15} can be carried out easily, one finds
\begin{align}
  t& =\frac{1}{6 \sqrt{3}} \log \left(\frac{\sqrt{3}+\sqrt{-s_1^{4}+4 s_1^2-1}}{2-s_1^2}\right)
-\frac{1}{3 \sqrt{3}} \log \left(\frac{\sqrt{3} s_1^2+\sqrt{-s_1^2+4 s_1^2-1}}{1-2{s_1}^2}\right) ;\label{6_25}\\
\vartheta_1& =-\frac{1}{2} \arccos \frac{2-s_1^2}{\sqrt{3}}-\frac{1}{2 \sqrt{3}} \log \left(\frac{\sqrt{3}+\sqrt{-s_1^{4}+4 s_1^2-1}}{2-8_1^2}\right) \label{6_26} \\
& \qquad +\frac{1}{\sqrt{3}} \log \left(\frac{\left.\sqrt{3}{s_1}^2+\sqrt{-s_1^{4}+4 s_1^2-1}\right)}{1-2 s_1^2}\right). \nonumber
\end{align}
The value of $s_1^2$ must be between $2-\sqrt{3}$ and $\frac{1}{2}$; both $t$ and $\vartheta_1$ vanish when $s_1^2=2-\sqrt{3}$. At the moment $t=0$, the triangle is isosceles, and the side lengths are
$$
s_1^2=2-\sqrt{3}, \;
s_2^2=s_3^2=\frac{1+\sqrt{3}}{2} .
$$
Over time, the values of $s_1$ and $s_2$ increase, and $s_3$ decreases; only after an infinitely long time do the three threads lie in a straight line, namely $s_1^2=s_3^2=\frac{1}{2}, s_{2}^2=$ 2, thread 2 is in the middle between threads 1 and 3. The orbits of all three threads are spirals that approach the circle of radius $\sqrt{\frac{1}{ 2}}$ asymptotically in time; moreover, the orbits of threads 1 and 3 approach the origin asymptotically. If $t$ is reasonably large, the paths of threads 1 and 3 lie very close to the circle of radius $\sqrt{\frac{1}{2}}$, diametrically opposite each other, with velocity $\frac{3\sqrt{2}}{2}$, while thread 2 is at the origin.

\section{Third case: $m_1 = 2 m_2 = - 2 m_3$}

Equations~\eqref{2_8} and~\eqref{2_9} can now be written
\begin{align}
 2 \varrho_1^2+\varrho_2^2-\varrho_3^2& =2 \lambda \label{7_1} \\
 s_1 s_2^2=s_{3}^2 . \label{7_2}
\end{align}
Here, $\lambda$ means an arbitrary constant, the unit of length having been appropriately imposed. From Eq.~\eqref{2_12}, it follows that
\begin{equation} \label{7_3}
  s_1^2=4 \varrho_1^2, \quad s_2^2=\varrho_2^2-\lambda, \quad s_3^2=\varrho_3^2+3 \lambda .
\end{equation}
Putting these equations into Eq.~\eqref{7_2} yields
\begin{equation} \label{7_4}
  \varrho_1\left(\varrho_2^2-\lambda\right)=\varrho_3^2+3 \lambda
\end{equation}
From Eqs.~\eqref{7_1} and~\eqref{7_4} follows by eliminating $\varrho_3$ that
\begin{equation} \label{7_5}
  \left(\varrho_1-1\right)\left(\varrho_2^2-2 \varrho_2-\lambda-2\right)=2 \lambda+2 .
\end{equation}

Assuming that $\lambda+1$ is non-zero, this equation yields
\begin{equation}\label{7_6}
  \varrho_2^2=\frac{2 \varrho_1^2+\lambda \varrho_1+\lambda}{\varrho_1-1}
\end{equation}
and then from Eq.~\eqref{7_1},
\begin{equation} \label{7_7}
  \varrho_3^2=\frac{2 \varrho_1^2-\lambda \varrho_1+3 \lambda}{\varrho_1-1} .
\end{equation}

If $\lambda=-1$, then the right-hand side of Eq.~\eqref{7_5} vanishes, and the equation can be satisfied by setting either one or both of the factors on the left equal to zero. The formulas that result when the second factor is allowed to vanish are given by the general equations~\eqref{7_6} and~\eqref{7_7}. Equating the first factor to zero gives
\begin{align*}
  \varrho_1&=1, & \varrho_2^2-\varrho_3^2&=4 \\
s_1&=2, & s_2&=s_3 .
\end{align*}
According to these equations, the triangle of the three vortex threads is isosceles and of fixed dimensions. We do not want to go into this case in more detail since later, in Sec.~\ref{sec:11}, we will integrate the differential equations for the motion of three vortex filaments in general, under the assumption that the triangle of the three filaments is at all times isosceles.

If we assume that both factors vanish at the same time, we get
\begin{align*}
\varrho_1^2 & =1, & \varrho_2^2&=3, & \varrho_3^2=7 \\
s_1      & =2, & s_2&=2, & s_3=2 ;
\end{align*}
the triangle of the three vortices is constantly equilateral and does not change in size. The motion consists of a triangle rotation around the center of mass at constant speed.

With an arbitrary value of $\lambda$ we get from Eq.~\eqref{2_11}, with the help of equations~\eqref{7_6} and~\eqref{7_7}
\begin{equation} \label{7_8}
\begin{split}
 \cos \left(\vartheta_3-\vartheta_1\right) & =\frac{3 \varrho_1^2-\lambda}{2 \varrho_1} \sqrt{\frac{\varrho_1-1}{2 \varrho_1^{3}-\lambda \varrho_1+3 \lambda}} ;\\
 \cos \left(\vartheta_1-\vartheta_2\right)&=-\frac{\varrho_1^2+\lambda}{2 \varrho_1} \sqrt{\frac{\varrho_1-1}{2 \varrho_1^2+\lambda \varrho_1+\lambda}} ,
 \end{split}
\end{equation}
and from here
\begin{equation} \label{7_9}
  \begin{split}
  \sin \left(\vartheta_3-\vartheta_1\right)& =\frac{1}{2 \varrho_1} \sqrt{\frac{-\varrho_1^{5}+9 \varrho_1^{4}+2 \lambda \varrho_1^{3}+6 \lambda \varrho_1^2-\lambda^2 \varrho_1+\lambda^2}{2 \varrho_1^{3}-\lambda \varrho_1+3 \lambda}}; \\ 
  \sin \left(\vartheta_1-\vartheta_2\right)&=-\frac{1}{2 \varrho_1} \sqrt{\frac{-\varrho_1^{5}+9 \varrho_1^{4}+2 \lambda \varrho_1^{3}+6 \lambda \varrho_1^{3}-\lambda^2 \varrho_1+\lambda^2}{2 \varrho_1^2+\lambda \varrho_1+\lambda} }.
  \end{split}
\end{equation}
The sign of one of the two sines can be chosen arbitrarily, and the sign of the other is then determined according to Eq.~\eqref{2_10}. Using the above, the following two equations result from the first of the differential equations~\eqref{7_3} and~\eqref{7_4}:
\begin{align}
 \dv{\varrho_1}{t}& =-\frac{\left(\varrho_1-1\right)^{3 / 2}\left(-\varrho_1^{5}+9 \varrho_1^{4}+2 \lambda \varrho_1^{3}+6 \lambda \varrho_1^2-\lambda^2 \varrho_1+\lambda^2\right)^{1 / 2}}{4 \varrho_1^2\left(\varrho_1^2+\lambda\right)} \label{7_10}\\
 \dv{\vartheta_1}{t}& =
  \frac{\left(\varrho_1-1\right)\left(\varrho_1^{3}+3 \varrho_1^2-\lambda \varrho_1+\lambda\right)}{4 \varrho_1^2\left(\varrho_1^2+\lambda\right)} . \label{7_11}
\end{align}
Eliminating $t$ from Eqs.~\eqref{7_10} and~\eqref{7_11} gives
\begin{equation} \label{7_12}
  \dv{\varrho_1}{ \vartheta_1}=\frac{-\varrho_1\left(\varrho_1-1\right)^{1 / 2}\left(-\varrho_1^{5}+9 \varrho_1^{4}+2 \lambda \varrho_1^{3}+6 \lambda \varrho_1^2-\lambda^2 \varrho_1+\lambda^2\right)^{1 / 2}}{\varrho_1^{3}+3 \varrho_1^2-\lambda \varrho_1+\lambda} .
\end{equation}

By quadrature one obtains $t$ and $\vartheta_1$ from Eqs.~\eqref{7_10} and~\eqref{7_12} through hyperelliptic integrals as functions of $\varrho_1$. For some values of $\lambda$, reductions occur. If $\lambda=-1$ or if it satisfies the equation
$$
\lambda^2-35 \lambda-243=0,
$$
then the integrals are elliptic. The simplest calculations are for $\lambda=0$, and we choose to continue for this value. From Eqs.~\eqref{7_10} and ~\eqref{7_12} we get for $\lambda=0$
\begin{align}
\dd t & =\frac{-4 \varrho_1^2 \dd \varrho_1}{\left(\varrho_1-1\right) \sqrt{\left(\varrho_1-1\right)\left(9-\varrho_1\right)}} ;\\
\dd \vartheta_1 & =\frac{-\left(\varrho_1+3\right) \dd \varrho_1}{\varrho_1 \sqrt{\left(\varrho_1-1\right)\left(9-\varrho_1\right)}},
\end{align}
and by integrating these equations
\begin{align}
  t&=\sqrt{\frac{9-\varrho_1}{\varrho_1-1}}+4 \sqrt{\left(\varrho_1-1\right)\left(9-\varrho_1\right)}+24 \arccos \frac{\varrho_1-5}{4} ; \label{7_15}\\
\vartheta_1&=\arccos \frac{\varrho_1^2-8 \varrho_1+9}{2 \varrho_1} . \label{7_16}
\end{align}
The constants of integration have been determined so that for $\varrho_1=9$, both $t$ and $\vartheta_1$ vanish.

From the moment $t=0$ onwards, $\varrho_1$ continuously decreases from 9 to $1$ and $\vartheta_1$ increases from 0 to $2 \pi$. Eq.~\eqref{7_16} represents the trajectory of thread 1. This trajectory is a fourth-order curve whose equation can be written in orthogonal coordinates as follows
\begin{equation} \label{7_17}
y_1^{4}+2\left(x_1^2+2 x_1-23\right) y_1^2+\left(x_1+3\right)^2\left(x_1-1\right)\left(x_1-9\right)=0.
\end{equation}
The $x$ axis is the axis of symmetry, and the point $(x_1,y_1)=(-3,0)$ is a self-crossing point. Let us set
$$
  x_1+3=\varrho \cos \vartheta, \;
  y=\varrho \sin \vartheta,
$$
i.e., by placing the self-crossing point at the origin of polar coordinates, we get from Eq.~\eqref{7_17}
\begin{equation} \label{7_18}
\varrho=4+8 \cos \vartheta .
\end{equation}
According to this equation, the trajectory of thread 1 is the pedal locus of a circle of radius 4 in relation to a point at a distance 8 from the center.

From Eq.~\eqref{7_16}, it follows that
$$
\begin{aligned}
 \cos \vartheta_1&=\frac{\varrho_1^2-8 \varrho_1+9}{2 \varrho_1} ;\\
 \sin \vartheta_1&=\frac{\varrho_1-3}{2 \varrho_1} \sqrt{\left(\varrho_1-1\right)\left(9-\varrho_1\right)},
\end{aligned}
$$
and also from Eqs.~\eqref{7_8} and~\eqref{7_9},
$$
\begin{aligned}
& \cos \left(\vartheta_1-\vartheta_2\right)=-\frac{1}{2} \sqrt{\frac{\varrho_1-1}{2}} ; \\
& \sin \left(\vartheta_1-\vartheta_2\right)=-\frac{1}{2} \sqrt{\frac{9-\varrho_1}{2}} .
\end{aligned}
$$
Substituting these expressions into the trigonometric identity
$$
\cos \vartheta_2=\cos \vartheta_1 \cos \left(\vartheta_1-\vartheta_2\right)+\sin \vartheta_1 \sin \left(\vartheta_1-\vartheta_2\right),
$$
one obtains
$$
\cos \vartheta_2=-\frac{2 \varrho_1-9}{2 \varrho_1} \sqrt{\frac{\varrho_1-1}{2}},
$$
or
$$
x_2=-\frac{2 \varrho_1-9}{2} .
$$
From Eq.~\eqref{7_6}, it follows for $\lambda=0$ that
$$
\varrho_1=\frac{\varrho_2^2-\sqrt{\varrho_2^{4}-8 \varrho_2^2}}{4} .
$$
By introducing this expression for $\varrho_1$ into the previous equation and converting this to rectangular coordinates, one obtains the equation of the path of thread 2, which can be written in the following form
\begin{equation} \label{7_19}
  \left(2 x_2-7\right) y_2^2+\left(x_2-3\right)^2\left(2 x_2+9\right)=0 .
\end{equation}
This equation represents a third-order curve, for which the $x$-axis is an axis of symmetry, the point $x_2=3, y_2=0$ is a self-crossing point, the line $2 x_2-7= 0$ is the unique real asymptote. Real values arise for $y_2$ only if $-\frac{9}{2} < x_2 < \frac{7}{2}$.

In the same way, as we determined the trajectory of thread 2, the trajectory of thread 3 follows. One first finds
$$
x_3=\varrho_1^2-9 \varrho_1+\frac{27}{2},
$$
and from here
$$
\varrho_1=\frac{9+\sqrt{27+4 x_3}}{2} .
$$
Substituting this expression for $\varrho_1$ in Eq.~\eqref{7_7} and then introducing rectangular coordinates, the following equation 
\begin{equation} \label{7_20}
  \left(2 x_3-11\right) y_{3}^{4}+2\left(x_3-9\right)\left(2 x_3^{3}+3 x_{3}-81\right) y_{3}^2 +\left(2 x_{3}-27\right)\left(x_3^2+2 x_{3}-27\right)^2=0.
\end{equation}
The trajectory of thread 3 is a fifth-order curve, for which the $x$-axis is the symmetry axis, the straight line $2 x_3-11=0$ is the only real asymptote, and the curve crosses itself at the points
$$
x_3=-1 \pm 2 \sqrt{7}, y_3=0.
$$
The coordinate $y_3$ only takes real values for values of $x_3$ between $-\frac{27}{4}$ and $\frac{27}{2}$. In particular, four values are real, if $x_3$ is between $-\frac{27}{4}$ and $\frac{11}{2}$, whereas only two if $x_3$ is between $\frac{11 }{2}$ and $\frac{27}{2}$ is located. The fifth-order curve still has two real self-crossing points. To find them, solve Eq.~\eqref{7_20} for $y_3^2$, which yields
\begin{equation}\label{7_21}
  y_3^2=\frac{-\left(x_3-9\right)\left(2 x_3^2+3 x_3-81\right) \pm 4\left(5 x_3-27\right) \sqrt{4 x_3+27}}{2 x_3-11} .
\end{equation}
The two values of $y_3^2$ corresponding to each $x_3$ coincide when either $5 x_3-27$ or $4 x_3+27$ vanish. The two self-crossing points, which correspond to the disappearance of the first quantity, are thus
$$
x_3=\frac{27}{5}, y_3= \pm \frac{54}{5} .
$$
$x_3=-\frac{27}{4}$ is one of the limits given above for $x_3$, and $4 x_3+27=0$ is the equation of a double tangent.

We still want to determine the speeds at which the threads move. Eqs~\eqref{7_10} and~\eqref{7_12} result, for $\lambda=0$, 
\begin{equation} \label{7_22}
\begin{split}
 \dv{\varrho_1}{t}& =-\frac{\varrho_1-1}{4 \varrho_1^2} \sqrt{\left(\varrho_1-1\right)\left(9-\varrho_1\right)} ;\\
 \dv{\vartheta_1}{t} &=\frac{\left(\varrho_1-1\right)\left(\varrho_1+3\right)}{4 \varrho_1^{3}},
\end{split}
\end{equation}
and it follows from this that
\begin{equation}\label{7_23}
  w_1=\frac{\varrho_1-1}{\varrho_1 \sqrt{\varrho_1}} .
\end{equation}
From the differential equations~\eqref{2_3} and~\eqref{2_4} one also obtains
\begin{align} 
  \dv{\varrho_2}{t}&=-\frac{\varrho_1-2}{4 \varrho_1^{3}} \sqrt{\frac{9-\varrho_1}{2}}, &\dv{\vartheta_2}{t}&=\frac{\left(\varrho_1-1\right)\left(\varrho_1+6\right)}{2 \varrho_1^{3}}, \label{7_24} \\
  \dv{\varrho_3}{t}&=-\frac{2 \varrho_1-3}{4 \varrho_1} \sqrt{\frac{9-\varrho_1}{2 \varrho_1}},& \dv{\vartheta_3}{t}&=\frac{\left(\varrho_1-1\right)\left(2 \varrho_1+9\right)}{2 \varrho_1^{3},} \label{7_25}
\end{align}
and from these
\begin{align}
 w_2&=\frac{1}{2 \varrho_1} \sqrt{\frac{3 \varrho_1-2}{\varrho_1}} \label{7_26}; \\
 w_3&=\frac{\sqrt{10 \varrho_1-9}}{2 \varrho_1} .
 \label{7_27}
\end{align}
For $\varrho_1=3$, $w_1=\frac{2}{9} \sqrt{3}$ is a maximum, for $\varrho_1=$ $\frac{9}{ 5}, w_3=\frac{5}{6}$ also a maximum. $w_2$ keeps growing from $t$ $=0$ onwards. We are now able to describe the motion fully.

At time $t=0$ all three threads are on the $x$-axis, with
$$
x_1=9, \; x_2=-\frac{9}{2},\;  x_3=\frac{27}{2} .
$$
The speeds are at this moment
$$
w_1=\frac{8}{27}, \; w_2=\frac{5}{54}, \; w_3=\frac{1}{2} .
$$
Thread 1 moves counterclockwise around the center of gravity and approaches it more and more, at first with increasing speed. After half a revolution, when $x_1=-3$ and $y_1=0$, the speed has reached the maximum, and from now on, it decreases constantly so that only after an infinitely long time does the thread in the location $x_1=1$, $y_1=0$. 

Thread 2 also goes around the center of gravity in a positive sense, with constantly increasing speed, only approaching it up to the distance $2 \sqrt{2}$; at this moment the triangle of the three threads is isosceles, $s_1=s_3=4, s_2=2 \sqrt{2}$; but then moves away more and more and finally moves very approximately with the speed $\frac{1}{2}$ on the straight line $x_2=\frac{7}{2}$ in the sense of the increasing $y $. 

\begin{figure}
  \centering
  \includegraphics[width=0.5\textwidth]{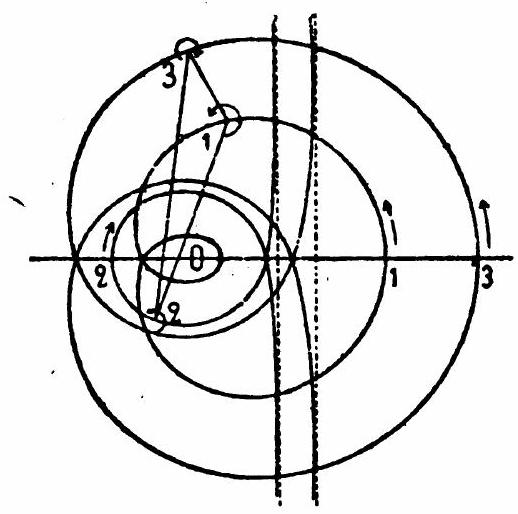}
  \caption{Gröbli's Figure 5.}
  \label{fig:5}
\end{figure}

Thread 3 also travels clockwise around the origin and initially approaches it with increasing speed. When $\varrho_3=\frac{27}{10} \sqrt{2}$ the velocity reaches the maximum value $\frac{5}{6}$ and from now on decreases and converges towards the limit $\frac{1}{2}$. The distance from the center of gravity becomes even smaller
and reaches the minimum, $\frac{3}{2} \sqrt{6}$, at a moment when the triangle is isosceles, $s_1=s_2=3$, $s_3 =\frac{3}{2} \sqrt{6}$, then increases indefinitely. Soon the motion is approximated that the thread follows the line $x_3=\frac{11}{2}$ with of velocity $\frac{1}{2}$. Then $y_3=y_2$. This is summarized in Fig.~\ref{fig:5}

\section{} 
So far, we have determined the motion of three vortex filaments for some particular value systems of the constants $m$. In doing so, we encountered particularly simple motions on various occasions, corresponding to certain specific solutions of the differential equations. We now set ourselves to look for some particular solutions of the differential equations, which determine the motion of three vortex filaments, without assigning specific values to the quantities $m$. We will make certain assumptions about the motion, investigate whether and, if so, under what conditions these assumptions are compatible with the differential equations, and, in the latter case, integrate the differential equations. First, we assume that the triangle of the three vortex threads changes neither shape nor size; then, we will assume that the triangle changes its size but not its shape; finally, we assume that the triangle is constantly isosceles.

\section{Triangles of vortex threads of unchanging shape or size}

For the right-hand sides of differential equations~\eqref{2_15} to vanish, either the triangle must be equilateral, or the three threads must lie in a straight line.

In the first case, by choosing the unit appropriate to the length, we may assume
\begin{equation} \label{9_1}
  s_1=s_2=s_3=1.
\end{equation}
From the differential equations~\eqref{2_17}, it results
\begin{equation} \label{9_2}
  \dv{\vartheta_1}{t}=\dv{\vartheta_2}{t}=\dv{\vartheta_3}{t}=\frac{m_1+m_2+m_3}{\pi} ;
\end{equation}
thus, the triangle of the three vortex threads rotates at a constant speed around the center of gravity. The radii of the circles in which the vortex threads move are respectively
\begin{equation} \label{9_3}
  \begin{split}
     \varrho_1 &= \frac{\sqrt{m_2^2+m_2 m_3+m_3^2}}{m_1+m_2+m_3} ;\\
     \varrho_2 &= \frac{\sqrt{m_3^2+m_3 m_1+m_1^2}}{m_1+m_2+m_3} ;\\
     \varrho_3 &= \frac{\sqrt{m_1^2+m_1 m_2+m_2^2}}{m_1+m_2+m_3} .
  \end{split}
\end{equation}
If $m_1+m_2+m_3=0$, i.e., when the center of gravity sits at infinity, then the three threads move in parallel straight lines, perpendicular to the direction of the center of gravity, with the velocity
$$
\frac{1}{\pi} \sqrt{\frac{m_1^2+m_2^2+m_3^2}{2}} .
$$

In the second case, it is advisable to use the differential equations~\eqref{2_3} and~\eqref{2_4}. If we place the origin at the center of gravity and also allow the quantities $\varrho$ to assume negative values, then we can set
\begin{equation} \label{9_4}
  \vartheta_1=\vartheta_2=\vartheta_3=\vartheta.
\end{equation}
Since the quantities $\varrho_1$, $\varrho_2$, and $\varrho_3$ are constant according to the assumption, the differential equations~\eqref{9_3} are fulfilled; Equations~\eqref{9_4} merge into the following
\begin{equation} \label{9_5}
  \begin{split}
    \pi \varrho_1 \dv{\vartheta}{t}&=\frac{m_2}{\varrho_1-\varrho_2}-\frac{m_3}{\varrho_3-\varrho_1} ; \\
    \pi \varrho_2 \dv{\vartheta}{t}&=\frac{m_3}{\varrho_2-\varrho_3}-\frac{m_1}{\varrho_1-\varrho_2} ; \\
    \pi \varrho_3 \dv{\vartheta}{t}&=\frac{m_1}{\varrho_3-\varrho_1}-\frac{m_2}{\varrho_2-\varrho_3} .
  \end{split}
\end{equation}
We multiply these equations in sequence by the following groups of factors
$$
\begin{aligned}
 m_1,\ & \frac{1}{\varrho_2-\varrho_3}, & m_1 \varrho_1; \\
 m_2,\ & \frac{1}{\varrho_3-\varrho_1}, & m_2 \varrho_2 ;\\
 m_3,\ & \frac{1}{\varrho_1-\varrho_2}, & m_3 \varrho_3,
\end{aligned}
$$
and add each time. In this way, the following equations are obtained, which replace the previous ones in all cases
\begin{equation} \label{9_6}
  \begin{split}
    0 & =m_1 \varrho_1+m_2 \varrho_2+m_3 \varrho_3; \\
    0 & =\frac{\varrho_1}{\varrho_2-\varrho_3}+\frac{\varrho_2}{\varrho_3-\varrho_1}+\frac{\varrho_3}{\varrho_1-\varrho_2} ;\\
\dv{\vartheta}{t} & =\frac{1}{\pi} \frac{m_2 m_3+m_3 m_1+m_1 m_2}{m_1 \varrho_1^2+m_2 \varrho_2^2+m_3 \varrho_3^2} .
  \end{split}
\end{equation}

The first two equations determine the ratios of the $\varrho$ variables and the third results in the angular velocity. Every set of $m$ values corresponds to three sets of ratios of $\varrho$ values, of which at least one is real. To every set of $\varrho$ values, which according to the second equation of~\eqref{9_6} must not be assumed arbitrarily, there belongs an infinite number of relationships between the values $m_1$, $m_2$, and $m_3$. A few special cases should also be highlighted.

\paragraph{1)} Let two of the quantities $m$, say $m_2$ and $m_3$, be equal to each other. Then, the first two equations of~\eqref{9_6} are satisfied by
$$
\varrho_1=0, \varrho_2+\varrho_3=0.
$$
There are also two other sets of values for the ratios of the $\varrho$, but these need not be real-valued.

\paragraph{2)} Let the sum of two of the constants $m$ equal zero, e.g., $m_2+m_3=0$. Then, the above equations are satisfied if
$$
\varrho_1=0, \varrho_2=\varrho_3 ;
$$
in this case, however, only one vortex thread remains. There are two additional sets of values for the ratios of $\varrho_1$, $\varrho_2$, and $\varrho_3$, which can be real or imaginary, depending on the values of the  $m$ parameters.

\paragraph{3)} Consider the case
\begin{equation} \label{9_7}
  m_2 m_3+m_3 m_1+m_1 m_2=0 .
\end{equation}
A solution of system~\eqref{9_6} is
\begin{equation} \label{9_8}
  \begin{gathered}
    \varrho_1: \varrho_2: \varrho_3=\frac{m_2-m_3}{m_1} : \frac{m_3-m_1}{m_2} : \frac{m_1-m_2}{m_3} \\
\dv{\vartheta}{t} =0,
  \end{gathered}
\end{equation}
and the three threads remain at rest. The other two sets of $\varrho$ values are solutions if they satisfy the equations
\begin{equation} \label{9_9}
  \begin{split}
     m_1 \varrho_1+m_2 \varrho_2+m_3 \varrho_3 &=0 \\
     m_1 \varrho_1^2+m_2 \varrho_2^2+m_3 \varrho_3^2 &=0 .
  \end{split}
\end{equation}
From these, we get if we denote an arbitrary constant by $x$
\begin{equation} \label{9_10}
  \begin{split}
x \varrho_1&=m_2 m_3\left(2 m_1-m_2-m_3\right)+\left(m_2-m_3\right) \sqrt{-m_1 m_2 m_3\left(m_1+m_2+m_{3}\right)} \\
x \varrho_2&=m_3 m_1\left(2 m_2-m_3-m_1\right)+\left(m_3-m_1\right) \sqrt{-m_1 m_2 m_3\left(m_1+m_2+m_3\right)} \\
x \varrho_3&=m_1 m_2\left(2 m_3-m_1-m_2\right)+\left(m_1-m_2\right) \sqrt{-m_1 m_2 m_3\left(m_1+m_2+m_3\right)} .
  \end{split}
\end{equation}
The right-hand side of the third equation of~\eqref{9_6} appears in the indefinite form $\frac{0}{0}$. One obtains the true value of the angular velocity from any equation from system~\eqref{9_5}. By combining these, one can easily create symmetric expressions for $\dv{\vartheta}{t}$; such an expression is, e.g.,
\begin{equation} \label{9_11}
  \dv{\vartheta}{t}=-\frac{m_2 m_3 \varrho_1+m_3 m_1 \varrho_2+m_1 m_2 \varrho_3}{\pi\left(m_1+m_2+m_3\right) \varrho_1 \varrho_2 \varrho_3}
\end{equation}

\paragraph{4)} Consider the case
\begin{equation} \label{9_12}
  m_1+m_2+m_3=0 .
\end{equation}
The first two of the equations~\eqref{9_6} are satisfied for $\varrho_1=\varrho_2=\varrho_3$ and this solution, which has no meaning for our problem, has to be counted twice. From the first equation~\eqref{9_6} 
\begin{equation} \label{9_13}
  \frac{\varrho_2-\varrho_3}{m_1}=\frac{\varrho_3-\varrho_1}{m_2}=\frac{\varrho_1-\varrho_2}{m_3}
\end{equation}
and now the second of these equations becomes
\begin{equation} \label{9_14}
  \frac{\varrho_1}{m_1}+\frac{\varrho_2}{m_2}+\frac{\varrho_3}{m_3}=0
\end{equation}
From~\eqref{9_14} and the first equation of~\eqref{9_6}, it now follows that for any arbitrary constant $x$,
\begin{equation} \label{9_15}
  \begin{split}
\varrho_1&=\frac{m_2^2-m_3^2}{m_2 m_3} x ;\\
\varrho_2&=\frac{m_3^2-m_1^2}{m_3 m_1} x ;\\
\varrho_3&=\frac{m_1^2-m_2^2}{m_1 m_2} x .
  \end{split}
\end{equation}
The angular velocity is given by the equation
\begin{equation}
 \dv{\vartheta}{t}=-\frac{1}{\pi x^2} \frac{m_1 m_2 m_3}{m_2 m_{3 }+m_3 m_1+m_1 m_2} .
\end{equation}

\section{The triangle of the three vortices that changes its size but not its shape}

From equations~\eqref{2_15}, it follows that the following derivatives must be constant
\begin{equation*} 
  \dv{\left(s_1^2\right)}{t}, \dv{\left(s_2^2\right)}{t}, \dv{\left(s_3^2\right)}{t}.
\end{equation*}
We have just dealt with the case that these derivatives disappear, so the triangle's size also remains fixed. If we use $\lambda_1$, $\lambda_2$, $\lambda_3$ to denote certain constants that still need to be determined and we have the starting point of time, then we can write
\begin{equation} \label{10_1}
  s_1^2=\lambda_1 t, \quad{s_2}^2=\lambda_2 t, \quad s_3^2=\lambda_3 t .
\end{equation}
If we insert these expressions into the differential equations~\eqref{2_15} and divide them one after the other by $\lambda_1$, $\lambda_2$, $\lambda_3$, then the left-hand sides are all equal to 1; by comparing the right-hand sides, it follows that
\begin{equation} \label{10_2}
m_1\left(\lambda_2-\lambda_3\right)=m_2\left(\lambda_3-\lambda_1\right)=m_3\left(\lambda_1-\lambda_2\right) .
\end{equation}
This double equation can be represented by the three equations
\begin{equation} \label{10_3}
  \begin{split}
     \lambda_2-\lambda_3&=\frac{m_1+m_2+m_3}{m_1} \mu; \\
     \lambda_3-\lambda_1&=\frac{m_1+m_2+m_3}{m_2} \mu; \\
     \lambda_1-\lambda_2&=\frac{m_1+m_2+m_3}{m_3} \mu,
  \end{split}
\end{equation}
where $\mu$ means an arbitrary constant. Adding these equations results in the condition between the variables $m$
\begin{equation} \label{10_4}
  \frac{1}{m_1}+\frac{1}{m_2}+\frac{1}{m_3}=0,
\end{equation}
which also emerges directly from the equation
$$
\frac{1}{m_1} \log s_1+\frac{1}{m_2} \log s_2+\frac{1}{m_3} \log s_3=\text { const. }
$$
Instead of the quantities $\lambda_1$, $\lambda_2$, and $\lambda_3$ we introduce three new quantities $\mu_1$, $\mu_2$, $\mu_3$ through the equations
\begin{equation} \label{10_5}
  \lambda_1=\mu \mu_1, \quad \lambda_2=\mu \mu_2, \quad \lambda_3=\mu \mu_3
\end{equation}
Equations~\eqref{10_3} thus simplify into the following 
\begin{equation} \label{10_6}
  \begin{split}
    \mu_2-\mu_3 &= \frac{m_1+m_2+m_3}{m_1}; \\
    \mu_3-\mu_1 &= \frac{m_1+m_2+m_3}{m_2}; \\
    \mu_1-\mu_2 &= \frac{m_1+m_2+m_3}{m_3} .
  \end{split}
\end{equation}
Provided that condition~\eqref{10_4} exists between the constants $m$, 
\begin{equation} \label{10_7}
  \begin{split}
    \frac{m_1+m_2+m_3}{m_1}&=-\frac{m_3-m_1}{m_3}+\frac{m_1-m_2}{m_3}; \\
    \frac{m_1+m_2+m_3}{m_2}&=-\frac{m_1-m_2}{m_3}+\frac{m_2-m_3}{m_1}; \\
    \frac{m_1+m_2+m_3}{m_3}&=-\frac{m_2-m_3}{m_1}+\frac{m_3-m_1}{m_2},
  \end{split}
\end{equation}
and from Eqs.~\eqref{10_6} and~\eqref{10_7} the correctness of the following equations becomes clear
\begin{equation} \label{10_8}
  \begin{split}
    \mu_1&=a-\frac{m_2-m_3}{m_1}; \\
    \mu_2&=a-\frac{m_3-m_1}{m_2}; \\
    \mu_3&=a-\frac{m_1-m_2}{m_3},
  \end{split}
\end{equation}
for some arbitrary constant $a$. Since the quantities $\mu_1$, $\mu_2$, and $\mu_3$ all have the same sign, the constant $a$ must either be larger than the largest of the expressions
$$
\frac{m_2-m_3}{m_1}, \quad
\frac{m_3-m_1}{m_2}, \quad
\frac{m_1-m_2}{m_3},
$$
or smaller than the smallest of them. Making use of equations~\eqref{10_1},~\eqref{10_5}, and~\eqref{10_6}, any one of the equations in system~\eqref{2_15} results in
\begin{equation} \label{10_9}
  \mu=\frac{m_1+m_2+m_3}{\pi} \frac{\sqrt{2 \mu_2 \mu_3+2 \mu_3 \mu_1+2 \mu_1 \mu_2-\mu_1^2-\mu_2^2-\mu_3^2}}{\mu_1 \mu_3 \mu_3} .
\end{equation}

According to Eq.~\eqref{2_12} the quantities $\varrho$ and $s$ must then satisfy:
\begin{equation} \label{10_10}
  \begin{split}
 \left(m_1+m_2+m_3\right) \varrho_1^2&=\left(m_2+m_3\right) s_1^2 \\ \left(m_1+m_2+m_3\right) \varrho_2^2&=\left(m_3+m_1\right) s_2^2 \\
 \left(m_1+m_2+m_3\right) \varrho_3^2&=\left(m_1+m_2\right) s_3^2 .
  \end{split}
\end{equation}

System~\eqref{2_17} now results in
\begin{equation} \label{10_11}
  \dd \vartheta_1= \dd \vartheta_2 = \dd \vartheta_3=\frac{x}{2} \frac{\dd t}{t},
\end{equation}
where, for brevity, we have set
\begin{equation} \label{10_12}
  x=\frac{m_1+m_2+m_3}{\pi} \frac{2 a^2+\frac{\left(m_2-m_3\right)\left(m_3-m_1\right)\left(m_1-m_2\right)}{m_1 m_2 m_3} a-3}{\mu \mu_1 \mu_2 \mu_3}.
\end{equation}
Defining $\vartheta$ by the equation
\begin{equation} \label{10_13}
  \vartheta=\frac{x}{2} \log t,
\end{equation}
it follows from Eq.~\eqref{10_11} that
\begin{equation} \label{10_14}
  \vartheta_1=\vartheta+\alpha_1, \quad 
  \vartheta_2=\vartheta+\alpha_2, \quad 
  \vartheta_3=\vartheta+\alpha_3 .
\end{equation}
One of the three constants $\alpha_1$, $\alpha_2$, $\alpha_3$ can be assumed arbitrarily; the other two are then determined by Equations~\eqref{2_11} and by the condition that one has to go around the triangle of the three threads in a positive or negative sense to reach threads 1, 2, 3 one after the other, depending on whether one takes the root occurring in Eq.~\eqref{10_9} with a negative or positive sign.

From Eqs.~\eqref{10_1},~\eqref{10_10},~\eqref{10_13}, and~\eqref{10_14}, we find the equations of the paths traced by the threads:
\begin{equation} \label{10_15}
  \begin{split}
& \varrho_1=\sqrt{\frac{m_2+m_3}{m_1+m_2+m_3} \mu \mu_1} e^{\frac{\vartheta_1-\alpha_1}{x}}; \\
& \varrho_2=\sqrt{\frac{m_3+m_1}{m_1+m_2+m_3} \mu \mu_2} e^{\frac{\vartheta_2-\alpha_2}{x}}; \\
& \varrho_3=\sqrt{\frac{m_1+m_3}{m_1+m_2+m_3} \mu \mu_3} e^{\frac{\vartheta_3-\alpha_3}{x}} .
  \end{split}
\end{equation}
According to these equations, the orbits are logarithmic spirals, and all three can be made to coincide with each other by rotating around the starting point.

For given values of the quantities $m$, since $a$ means an arbitrary constant, infinitely many triangle shapes are possible. The triangle is right-angled if $a$ is equal to one of the values 
$$
-\frac{m_2-m_3}{m_2+m_3}, \quad
-\frac{m_3-m_1}{m_3+m_1}, \quad
-\frac{m_1-m_2}{m_1+m_2},
$$
two of which always satisfy the conditions to which $a$ is subject. The isosceles shape of the triangle is impossible.

\begin{figure}
  \centering
  \includegraphics[width=0.6\textwidth]{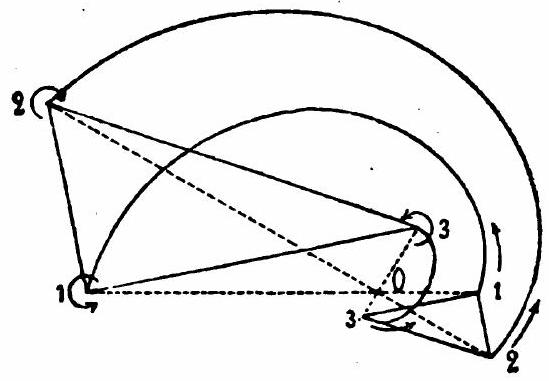}
  \caption{Gröbli's Figure 6}
  \label{fig:6}
\end{figure}

Figure~\ref{fig:6} corresponds to the choice
$$
\begin{gathered}
m_1: m_2: m_3=3:-2: 6; \\
a=2 .
\end{gathered}
$$
With a suitable choice of the time unit one obtains
$$
\begin{array}{lll}
s_1^2=28 t, & s_2^2=21 t, & s_3^2=7 t ; \\
\varrho_1^2=16 t, & \varrho_2^2=27 t, & \varrho_3^2=t ;
\end{array}
$$
$$
\begin{aligned}
\varrho_1&=4 e^{\frac{\sqrt{3}}{5}\left(\vartheta_1-\alpha_1\right)};\\
\varrho_2&=3 \sqrt{3} e^{\frac{\sqrt{3}}{5}\left(\vartheta_2-\alpha_2\right)};\\
\varrho_3&=e^{\frac{\sqrt{3}}{5}\left(\vartheta_3-\alpha_3\right)} ;
\end{aligned}
$$
$$
\begin{array}{lll}
\vartheta_2-\vartheta_3&=\alpha_2-\alpha_3&=\frac{\pi}{2};\\
\vartheta_3-\vartheta_1&=\alpha_3-\alpha_1&=-\frac{2 \pi}{3};\\
\vartheta_1-\vartheta_2&=\alpha_1-\alpha_2&=\frac{\pi}{6} 
\end{array}.
$$

\section{A constantly isosceles triangle of three vortices}
\label{sec:11}
Let's assume that $s_2=s_3$. From the first equation of~\eqref{2_15}, we get $\dv{\left(s_1^2\right)}{t}=0$. One may choose the unit of length so that
\begin{equation} \label{11_1}
  s_1 = 1.
\end{equation}
The left-hand sides of the second and third equations~\eqref{2_15} are equal to each other so that the right-hand sides are also equal, which requires.
\begin{equation} \label{11_2}
  m_2+m_3=0,
\end{equation}
This then implies
\begin{equation} \label{11_3}
  -\frac{m_2}{\pi} \dd t=\frac{s_2^2 \dd\left(s_2^2\right)}{\left(s_2^2-1\right) \sqrt{4 s_2^2-1}} .
\end{equation}
By integrating this equation and suitably determining the integration constants, it follows that
\begin{equation} \label{11_4}
\frac{2 m_2}{\pi} t
  = \begin{cases}
   -\sqrt{4 s_2^2-1}+\frac{2}{\sqrt{3}} \log \left(\frac{\sqrt{3}+\sqrt{4 s_2^2-1}}{\sqrt{3}-\sqrt{4 s_2^2-1}}\right), \frac{1}{2} \le s_2<1; \\
  -\sqrt{4 s_2^2-1}+\frac{2}{\sqrt{3}} \log \left(\frac{\sqrt{4 s_2^2-1}+\sqrt{3}}{\sqrt{4 s_2^2-1}-\sqrt{3}}\right), 1<s_2<\infty .
  \end{cases}
\end{equation}

From equations~\eqref{2_12}, one obtains
\begin{equation} \label{11_5}
  \begin{split}
  & \varrho_1^2=\frac{m_2^2}{m_1^2}; \\
  & \varrho_2^2=s_2^2+\frac{m_2\left(m_2-m_1\right)}{m_1^2} ;\\
  & \varrho_3^2=s_2^2+\frac{m_2\left(m_2+m_1\right)}{m_1^2} .
  \end{split}
\end{equation}
According to the first of these equations, thread 1 moves in a circle whose center coincides with the center of gravity of the vortex threads.

The first equation of~\eqref{2_17} turns into
$$
\dv{\vartheta_1}{t}=\frac{m_1}{\pi} \frac{1}{s_2^2} \text {. }
$$
Eliminating $t$ from this equation and from Eq.~\eqref{11_3} gives
\begin{equation} \label{11_6}
  \dd \vartheta_1=-\frac{m_1}{m_2} \dd{\left(s_2^2\right)}{\left(s_2^2-1\right) \sqrt{4 s_2^2-1}},
\end{equation}
and now by integration and appropriate choice of the $x$-axis
\begin{equation} \label{11_7}
  \vartheta_1 = \begin{cases}
\frac{m_1}{m_2 \sqrt{3}} \log \left(\frac{\sqrt{3}+\sqrt{4 s_2^2-1}}{\sqrt{3}-\sqrt{4 s_2^2-1}}\right), & \frac{1}{2} \le s_2<1 ; \\
\frac{m_1}{m_2 \sqrt{3}} \log \left(\frac{\sqrt{4 s_2^2-1}+\sqrt{3}}{\sqrt{4 s_2^2-1}-\sqrt{3}}\right), & 1<s_2<\infty .
\end{cases}
\end{equation}
Using the formulas~\eqref{2_11}, the equations for $\vartheta_2$ and $\vartheta_3$ become
\begin{equation} \label{11_8}
  \begin{split}
    & \vartheta_2=\vartheta_1-\arctan\left(\frac{m_1}{2 m_2-m_1} \sqrt{4 s_2^2-1}\right); \\
& \vartheta_3=\vartheta_1-\arctan\left(\frac{m_1}{2 m_2+m_1} \sqrt{4 s_2^2-1}\right) .
  \end{split}
\end{equation}
If one replaces $s_2$ by $\varrho_2$ or $\varrho_3$ using Equations~\eqref{11_5}, one obtains the equations for the paths traced by threads 2 and 3. These orbits are spirals which are asymptotic to the circles
\begin{equation} \label{11_9}
  \begin{split}
& \varrho_2^2=\frac{m_1^2-m_1 m_2+m_2^2}{m_2^2}; \\
& \varrho_3^2=\frac{m_1^2+m_1 m_2+m_2^2}{m_2^2}.
  \end{split}
\end{equation}
For $s_2>1$ the straight lines
\begin{equation} \label{11_10}
  \begin{split}
    & x_2=\frac{2 m_1^2-m_1 m_2+m_2^2}{m_1^2}; \\
& x_3=\frac{2 m_1^2+m_1 m_2+m_2^2}{m_2^2}
  \end{split}
\end{equation}
are also asymptotes. In the case $2 m_2=m_1$; we then have $\vartheta_2-\vartheta_1=-\frac{\pi}{2}$. The argument in Sec. 5 for $m_2=m_1$ should suffice for the rest.

\section{} 
In what has been said so far, we have twice found solutions of the differential equations for the motion of three vortex filaments corresponding to a motion of the vortex filaments along parallel straight lines. These were the cases discussed in Secs.~4 and~9
$$
m_1=m_2=-m_3 \qand m_1 \varrho_1^2+m_2 \varrho_2^2+m_3 \varrho_3^2=0
$$
and
$$
s_1=s_2=s_3 \qand m_1+m_2+m_3=0 .
$$

The task of determining all motions in which the paths are parallel straight lines can easily be done using the differential equations~\eqref{2_1}, and the answer is that these are the only two such solutions.

Likewise, the question of such solutions of the differential equations can be answered, in which the triangle of the three vortex filaments remains right-angled for all time. The two found in Secs.~4 and~10 are the only ones.

Here, we leave the problem of the movement of three vortex threads and proceed to determine the motions of four vortex threads, assuming a plane of symmetry.

\begin{center}
  \textbf{\LARGE The movement of four vortex threads, assuming a plane of symmetry.}
\end{center}

\section{} 
Assume there is an axis of symmetry for movement in the $x y$ plane. We take this to be the $y$-axis of a Cartesian coordinate system, arbitrarily assumed to contain the origin. Label the four threads 1, 2, 3, and 4, with threads 3 and 4  symmetrical to threads 1 and 2. Then
\begin{equation} \label{13_1}
\begin{split}
x_3=-x_1, \quad & x_{4}=-x_2 \\
y_3=y_1, \quad & y_{4}=y_2 .
\end{split}
\end{equation}
For the requirement that the $y$-axis should be the axis of symmetry of the movement to be compatible with the differential equations~\eqref{2_1}, the quantities $m$ defined by equation~\eqref{1_1} must meet the conditions
\begin{equation} \label{13_2}
  m_1+m_3=0, \quad m_2+m_{4}=0,
\end{equation}
and these conditions are also sufficient.

Of the four general integrals
$$
\begin{aligned}
\sum m_1 x_1 & =\mathrm { const. }, & \sum m_1 y_1 & =\mathrm { const. } \\
\sum m_1 \varrho_1^2 & =\mathrm { const. }, & P & =\mathrm { const. },
\end{aligned}
$$
the two in the middle become irrelevant because the left sides disappear identically. The first integral becomes
\begin{equation} \label{13_3}
  m_1 x_1+m_2 x_2=\mathrm{ const. }
\end{equation}
and states that the center of gravity of threads 1 and 2 moves parallel to the $y$-axis. Since everything that applies to threads 1 and 2 also applies to threads 3 and 4, we will only discuss threads 1 and 2 in the following. Using equations 
\begin{equation} \label{13_4}
\begin{split}
& \varrho_{12}^2=\varrho_{34}^2=\left(x_1-x_2\right)^2+\left(y_1-y_2\right)^2 ; \\
& \varrho_{14}^2=\varrho_{23}^2=\left(x_1+x_2\right)^2+\left(y_1-y_2\right)^2 ; \\
& \varrho_{13}^2=4 x_1^2, \quad \varrho_{24}^2=4 x_2^2
\end{split}
\end{equation}
The equation results from the last of the above integrals
\begin{equation} \label{13_5}
  \left\{\frac{\left(x_1-x_2\right)^2+\left(y_1-y_2\right)^2}{\left(x_1+x_2\right)^2+\left(y_1-y_2\right)^2}\right\}^{m_1 m_2} \frac{1}{x_1^{ m_1 m_1} x_2^{ m_2 m_2}}=\text { const. }
\end{equation}

The differential equations
$$
m_1 \dv{x_1}{t}=\frac{\partial P}{\partial y_1} \qand m_1 \dv{y_1}{t}=-\frac{\partial P}{\partial x_1}
$$
become
\begin{equation} \label{13_6}
  \begin{split}
& \dv{x_1}{t}=-\frac{m_2}{\pi}\left(y_1-y_2\right)\left(\frac{1}{\varrho_{12}^2}-\frac{1}{\varrho_{14}^2}\right) \\
& \dv{y_1}{t}=-\frac{m_2}{\pi}\left(\frac{x_1-x_2}{\varrho_{12}^2}-\frac{x_1+x_2}{\varrho_{14}^2}\right)-\frac{m_1}{\pi} \frac{1}{2 x_1} .
\end{split}
\end{equation}

Using equations~\eqref{13_3} and~\eqref{13_5}, the quantities $x_2$ and $y_1-y_2$ can be expressed in terms of $x_1$ and since, by system~\eqref{13_6}, the quantities $y_{1 }$ and $y_2$ appear only in the combination $y_1-y_2$, so the evolution equations take the form
\begin{equation} \label{13_7}
\dv{x_1}{t}=f_1\left(x_1\right), \quad \dv{y_1}{t}=f_2\left(x_1\right)  .
\end{equation}
From these, it follows by eliminating $t$ that
\begin{equation} \label{13_8}
  \dv{y_1}{x_1}=f_3\left(x_1\right).
\end{equation}

By integrating Eq.~\eqref{13_8} and the first equation of~\eqref{13_7}, we obtain $y_1$ and $t$ as functions of $x_1$. The movement of the first thread is thus determined, as is that of thread 2 according to Eqs.~\eqref{13_3} and~\eqref{13_5}.

The execution of all of these calculations in closed form is only possible for special values of the quantities $m_1$ and $m_2$. The simplest assumption one can make about $m_1$ and $m_2$ is
\begin{equation} \label{13_9}
  m_1=m_2
\end{equation}
We now wish to determine the corresponding movement.

\section{} 
For $m_1=m_2$, Eq.~\eqref{13_3} gives
$$
x_1+x_2=\text { const. }
$$
The constant is either zero or nonzero. In the latter case, a suitable choice of the unit of length can give it a simple value. If the constant is zero, then there is a second axis of symmetry parallel to the $x$-axis. Since we will later determine the general motion for $2 n$ vortex threads, assuming $n$ planes of symmetry, we can refrain from treating the case where the constant mentioned disappears. If we now give the constant the value 2, the above equation can be replaced by the two equations
\begin{equation} \label{14_10}
  x_1=1+x, \quad x_2=1-x,
\end{equation}
in which $x$ means the abscissa of 1 with respect to a coordinate system whose ordinate axis is the straight line in which the center of gravity of 1 and 2 moves. If $\lambda$ denotes an arbitrary, positive or negative constant, equation~\eqref{13_5} can be written
$$
\frac{\left(y_1-y_2\right)^2+4 x^2}{\left(y_1-y_2\right)^2+4} \frac{1}{1-x^2}=\frac{1}{\lambda};
$$
and from this, it follows
\begin{equation} \label{14_11}
  \begin{split}
    \left(y_1-y_2\right)^2 & =4 \frac{1-(\lambda+1) x^2}{\lambda-1+x^2} ;\\
\varrho_{12}^2=4 x^2+\left(y_1-y_2\right)^2 & =4 \frac{\left(1-x^2\right)^2}{\lambda-1+x^2} ;\\
\varrho_{14}^2=4+\left(y_1-y_2\right)^2 & =4 \frac{\lambda( 1-x^2)}{\lambda-1+x^2} .
  \end{split}
\end{equation}
We want to assume $m_1$ to be positive. By appropriately choosing the time unit, we can also give the quantity $m_1$ a simple value. We want to assume $m_1=m_2$ $=2 \pi$. The first equation of~\eqref{13_6} becomes, now that $\dd x_1=\dd x$
\begin{equation} \label{14_12}
\dv{x}{t} = - \frac{ \left(\lambda-1+x^2 \right)^\frac{3}{2} \left(1-(\lambda+1)x^2 \right)^\frac{1}{2}}{\lambda(1-x^2)^2},
\end{equation}
and from this, it follows that
\begin{equation} \label{14_13}
t=-\int \frac {\lambda(1-x^2)^2 \dd x}
{(\lambda-1+x^2) \sqrt{(\lambda-1+x^2)(1-(\lambda+1)x^2)}}.
\end{equation}
The second equation of system ~\eqref{13_6} becomes
\begin{equation} \label{14_14}
  \dv{y_1}{t} = \frac{x^4 + 2(\lambda-1)x^3 + \lambda^2 x + 1-2\lambda}
  {\lambda(1-x^2)^2}.
\end{equation}
Eliminating $t$ from Eqs.~\eqref{14_12} and~\eqref{14_14} we obtain
\begin{equation} \label{14_15}
  \dv{y_1}{x} = - \frac{x^4 + 2(\lambda-1)x^2 + \lambda^2 x + 1-2\lambda}
  { \left(\lambda-1+x^2 \right)^\frac{3}{2} \left(1-(\lambda+1)x^2 \right)^\frac{1}{2}}.
\end{equation}
By replacing $x$ with $-x$, one then finds
\begin{equation} \label{14_16}
  \dv{y_2}{x} = - \frac{x^4 + 2(\lambda-1)x^2 - \lambda^2 x + 1-2\lambda}
  { \left(\lambda-1+x^2 \right)^\frac{3}{2} \left(1-(\lambda+1)x^2 \right)^\frac{1}{2}}
\end{equation}
and it then follows from Eqs.~\eqref{14_15} and~\eqref{14_16} that
\begin{equation} \label{14_17}
\frac{y_1+y_2}{2} = - \int \frac{(x^4 + 2(\lambda-1)x^3 + 1-2\lambda) \dd x}
{ \left(\lambda-1+x^2 \right)^\frac{3}{2} \left(1-(\lambda+1)x^2 \right)^\frac{1}{2}}.
\end{equation}
From the first equation of System~\eqref{14_11}, we obtain
\begin{equation} \label{14_18}
  \frac{y_1-y_2}{2} = \sqrt{ \frac{1-(\lambda+1)x^2}{\lambda-1+x^2}}.
\end{equation}
So we have obtained $y_1$ and $y_2$ as functions of $x$, and according to Eq.~\eqref{14_10}, as functions of $x_1$ and $x_2$ respectively.

The integrals in Eqs.~\eqref{14_13} and~\eqref{14_17} are generally elliptic. For the further calculation, one has to distinguish between the following four cases  
\begin{equation*}
 \infty>\lambda>1, \quad 1>\lambda>0, \quad
 0>\lambda>-1, \quad -1>\lambda>-\infty .
\end{equation*}

The borderline cases are $\lambda=\infty, \lambda=0, \lambda=1, \lambda=-1$. We can ignore the first two because they lead back to two vortex threads. In the other two cases, the integrals in Eqs.~\eqref{14_13} and~\eqref{14_17} are logarithmic and algebraic.

\section{Case I: $\infty>\lambda>1$}
In order for the fourth degree function of $x$ under the root sign in Eqs.~\eqref{14_13} and~\eqref{14_17} to be positive, $x$ must satisfy the condition
\begin{equation*}
  \sqrt{\frac{1}{1+\lambda}} \ge x \ge -\sqrt{\frac{1}{1+\lambda}}.
\end{equation*}
When the elliptic integrals are reduced to Legendre's normal integrals, $\frac{1}{\lambda}$ results as a module. To stick with the usual designation, we put
\begin{equation} \label{15_19}
  \lambda=\frac{1}{\kappa} .
\end{equation}
The previous condition for $x$ then becomes
\begin{equation} \label{15_20}
  \sqrt{\frac{\kappa}{1+\kappa}} \ge x \ge -\sqrt{\frac{\kappa}{1+\kappa}} .
\end{equation}
Let us set
\begin{equation} \label{15_21}
  x=\sqrt{\frac{\kappa}{1+\kappa}} \cos{\psi},
\end{equation}
so the values $\psi=0$ and $\psi=\pi$ correspond to the limits of $x$ and it becomes
\begin{equation*}
  \frac{\dd x}{\sqrt{\left(\lambda-1+x^2\right)\left(1-(\lambda+1) x^2\right)}}=-\frac{\kappa \dd \psi}{\sqrt{1-\kappa^2 \sin^2{\psi}}} .
\end{equation*}
After carrying out some simple calculations, the following equations result from Eqs.~\eqref{14_13},~\eqref{14_17}, and~\eqref{14_18}, if the integration constants are determined appropriately
\begin{align}
t&=\frac{2}{\kappa\left(1-\kappa^2\right)} E(\kappa, \psi)-\frac{2}{\kappa} F(\kappa, \psi)-\frac{\kappa}{1-\kappa} \frac{\sin{\psi} \cos{\psi}}{\sqrt{1-\kappa^2 \sin^2{\psi}}} \label{15_22};\\
y_1&=-\frac{2 \kappa}{1-\kappa^2} E(\kappa, \psi)+\left(\frac{\kappa^2}{1-\kappa^2} \cos{\psi}+\sqrt{\kappa(1+\kappa)}\right) \frac{\sin{\psi}}{\sqrt{1-\kappa^2 \sin^2{\psi}}} \label{15_23};\\
y_2&=-\frac{2 \kappa}{1-\kappa^2} E(\kappa, \psi)+\left(\frac{\kappa^2}{1-\kappa^2} \cos{\psi}-\sqrt{\kappa(1+\kappa)}\right) \frac{\sin{\psi}}{\sqrt{1-\kappa^2 \sin^2{\psi}}}, \label{15_24}
\end{align}
in which $F(\kappa, \psi)$ and $E(\kappa, \psi)$ mean the Legendre integrals of the first and second kind. If we also apply substitution~\eqref{15_21} to Eqs.~\eqref{14_12},~\eqref{14_14} and the equation for $\dv{y_2}{t}$ formed accordingly, we get the equations
\begin{align}
\dv{x}{t} & =\dv{x_1}{t}=-\dv{x_2}{t}=-\sqrt{\frac{1+\kappa}{\kappa}} \frac{\sin{\psi}\left(1-\kappa^2 \sin^2{\psi}\right)^{3 / 2}}{\left(1+\kappa \sin^2{\psi}\right)^2}; \label{15_25} \\
\dv{y_1}{t} & =-\frac{2+\kappa+2 \kappa \sin^2{\psi}-\kappa^3 \sin^4{\psi}-(1+\kappa) \sqrt{\frac{1+\kappa}{\kappa}} \cos{\psi}}{\left(1+\kappa \sin^2{\psi}\right)^2}; \label{15_26}\\
\dv{y_2}{t} & =-\frac{2+\kappa+2 \kappa \sin^2{\psi}-\kappa^3 \sin^4{\psi}+(1+\kappa) \sqrt{\frac{1+\kappa}{\kappa}} \cos{\psi}}{\left(1+\kappa \sin^2{\psi}\right)^2} . \label{15_27}
\end{align}

These solutions are periodic. Replacing $\psi$ instead of $\psi+2 \pi$ leaves the equations for $x$ and the velocities unchanged, and additional terms are added to $t, y_1$ and $y_2$. If $K$ and $E$ denote the complete elliptic integrals of the first and second kind, then $t$ increases by
\begin{equation} \label{15_28}
  T=\frac{8}{\kappa}\left(\frac{1}{1-\kappa^2} E-K\right) .
\end{equation}
per period, and $y_1$ and $y_2$ decrease by
\begin{equation} \label{15_29}
  Y=\frac{8 \kappa}{1-\kappa^2} E.
\end{equation}
Here, $T$ is the duration of a period, and $Y$ is the distance by which the threads have shifted in the direction of the negative $y$ axis during this time. If you put $\psi+\pi$ in place of $\psi$, then $x_1$ becomes $x_2$, $\dv{x_1}{t}$ becomes $\frac{ d x_2}{t}$, $\dv{y_1}{t}$ becomes $\dv{y_2}{t}$ and vice versa,  and $t$ is increased by $\frac{1}{2} T$.

The path of thread 1 is a periodic curve in the direction of the $y$ axis whose period equals $Y$. The trajectory of 2 is the same curve, just shifted by $\frac{1}{2} Y$.

At time $t=0$, thread 1 is at point $x_1=1+\sqrt{\frac{1+\kappa}{\kappa}}$, $y_1=0$; its velocity is parallel to the $y$-axis. At the same moment, thread 2 is at the position $x_2=1-\sqrt{\frac{\kappa}{1+\kappa}}$, $y_2=0$; its velocity is also parallel to the $y$-axis. The coordinate $x_1$ now decreases while $y_1$ may initially increase or decrease; we leave it undecided, and $x_2$ and $y_2$ decrease. At time $t=\frac{1}{4} T$ both threads pass the line $x_1=x_2=1$ at different points, but with the same speed, and then continue with swapped speeds until time $t=\frac{1}{2} T$, at which point $y_1=y_2=-\frac{1}{2} Y$, $x_1=1-\sqrt{\frac {\kappa}{1+\kappa}}$, $x_2=1+\sqrt{\frac{\kappa}{1+\kappa}}$, etc.

To see more clearly the shape of the curves described by the vortex threads, we examine the behavior of the time derivatives $\dv{y_1}{t}$ and $\dv{y_2}{t}$. According to the above, we can limit ourselves to the values of $\psi$ between 0 and $\frac{\pi}{2}$. From Eq.~\eqref{15_27}, one can immediately see that in this interval $\dv{y_2}{t}$ is consistently negative. For $\psi=\frac{\pi}{2}$ $\dv{y_1}{t}$ is also negative, for $\psi=0$ this results
$$
\dv{y_1}{t}=-(2+\kappa)+(1+\kappa) \sqrt{\frac{1+\kappa}{\kappa}}
$$
and this expression can be positive or negative. To separate the two cases, we determine the value of $\kappa$ for which it vanishes. The resulting equation is
$$
\kappa^2+\kappa-1=0,
$$
and from this, the one usable root is 
$$
\kappa=\frac{\sqrt{5}-1}{2}=0.618 \ldots
$$
If $\kappa>\frac{\sqrt{5}-1}{2}$ then it satisfies
$$
2+\kappa>(1+\kappa) \sqrt{\frac{1+\kappa}{\kappa}}>(1+\kappa) \sqrt{\frac{1+\kappa}{\kappa}} \cos{\psi},
$$
and there
$$
2 \kappa \sin^2{\psi}-\kappa^{3} \sin^{4} \psi>0,
$$
so it follows that $\dv{y_1}{t}$ is consistently negative. In this case, $y_1$ and $y_2$ continue to decrease. If $\kappa<\frac{\sqrt{5}-1}{2}$, then $\dv{y_1}{t}$ vanishes between $\psi=0$ and $\psi=\frac{\pi}{2}$ at least once; the associated value of $\psi$ is determined from the equation
$$
2+\kappa+2 \kappa \sin^2{\psi}-\kappa^{3} \sin^{4} \psi=(1+\kappa) \sqrt{\frac{1+\kappa}{\kappa}} \cos{\psi} .
$$
As $\psi$ increases from 0 to $\frac{\pi}{2}$, the left side of this equation increases, and the right side decreases. The equation, therefore, has at most one root. This corresponds to a maximum of $y_1$, and the path has a self-crossing point. In the limit case $\kappa=\frac{\sqrt{5}-1}{2}$ this turns into a corner.

We now determine the inflect points. Using Eq.~\eqref{14_15},
$$
\dv{y_1}{x}=-\frac{x^{4}+2(\lambda-1) x^2+\lambda^2 x+1-2 \lambda} {\left(\lambda-1+x^2\right)^{3 / 2}\left(1-(\lambda+1) x^2\right)^{1 / 2}},
$$
the condition for the inflection points, $\dv[2]{y_1}{x}=0$, leads to a fifth-degree equation with one root  $x =1$. As can be seen from Eq.~\eqref{14_18}, $x$ can never take the value 1; after division by $x-1$, the equation remains fourth degree
\begin{equation} \label{15_30}
  f(x) \equiv x^{4}+(4+3 \lambda) x^{3}-(2-\lambda) x^2-(4-\lambda) x+1 -\lambda=0
\end{equation}
To investigate the reality of the roots of this equation, we need to determine the values of $\lambda$ for which two of them coincide. For this to be the case, the equation 
$$
f^{\prime}(x) \equiv 4 x^{3}+(12+9 \lambda) x^2-(4-2 \lambda) x-4+\lambda=0
$$
must also be satisfied. Eliminating $x$ from this and the previous equation results in a sixth-degree equation in $\lambda$. It is easier to eliminate $\lambda$, which yield the equation
$$
3 x^{6}+2 x^{5}+13 x^{4}+28 x^{3}-19 x^2+2 x+3=0,
$$
which has two real roots
\begin{equation} \label{15_31}
  x^{\prime}=-0.30186 \qand x^{\prime \prime}=-1.90134 .
\end{equation}
The corresponding values of $\lambda$ are
\begin{equation}
  \lambda^{\prime}=1.48732 \qand \lambda^{\prime \prime}=-0.65555 .
\end{equation}
The fourth-degree equation that determines the inflection points now has
\begin{align*}
\text{two real roots if } \infty                   &>\lambda  >\lambda^{\prime}, \\
\text{four real roots if } \lambda^{\prime}        &>\lambda > \lambda^{\prime \prime}, \\
\text{two real roots if } \lambda^{\prime \prime}  &>\lambda > - \infty.
\end{align*}
Note that the roots of this equation are the abscissae of the intersection points of the two curves
$$
\begin{aligned}
& y=\left(x^2-1\right)\left(x^2+4 x-1\right) \qand \\
& y=-\lambda\left(3 x^{3}+x^2+x+1\right).
\end{aligned}
$$
From this, one can easily arrive at boundaries within which the roots in the various cases lie. From Eq.~\eqref{15_30}, one finds
\begin{gather*}
f\left(\frac{1}{\sqrt{1+\lambda}}\right)=-\left(\frac{\lambda}{1+\lambda}\right)^2(\lambda -\sqrt{1+\lambda}) ; \\
f(\sqrt{1-\lambda})=-3 \lambda^2 \sqrt{1-\lambda}; \\
f(0)=1-\lambda; \quad f(\infty)=+\infty .
\end{gather*}

Using these results, you can now determine the inflection points of the curve traced by thread 1 and the path of thread 2 if you replace $x$ with $-x$.

In the present case it results that between $\psi=0$ and $\psi=\pi$ the curve described by thread 1 has
\begin{align*}
\text{no real inflection points for } &0<x<\frac{\sqrt{5}-1}{2}, \\
\text{one real inflection point for } &\frac{\sqrt{5}-1}{2}<x<\frac{1}{\lambda^{\prime}},\\
\text{three real inflection points for }& \frac{1}{\lambda^{\prime}}<x<1.
\end{align*}

The expressions
$$
T=\frac{8}{\kappa}\left(\frac{E}{1-\kappa^2} -K\right), 
Y=\frac{8 \kappa  E}{1-\kappa^2}, \text{ and } X=2 \sqrt{\frac{\kappa}{1+\kappa}},
$$
the last of which indicates the excursion that the threads make in the direction of the $x$-axis, obtain the following values for $\kappa=1$
$$
T=\infty, Y=\infty, \qand X=\sqrt{2} .
$$
If $\kappa$ decreases, all three quantities decrease and converge with $\kappa$ towards zero, and the latter two in such a way that the quotient $\frac{Y}{X}$ also becomes smaller and smaller.

\begin{figure}
  \centering
  \includegraphics[width=0.9\textwidth]{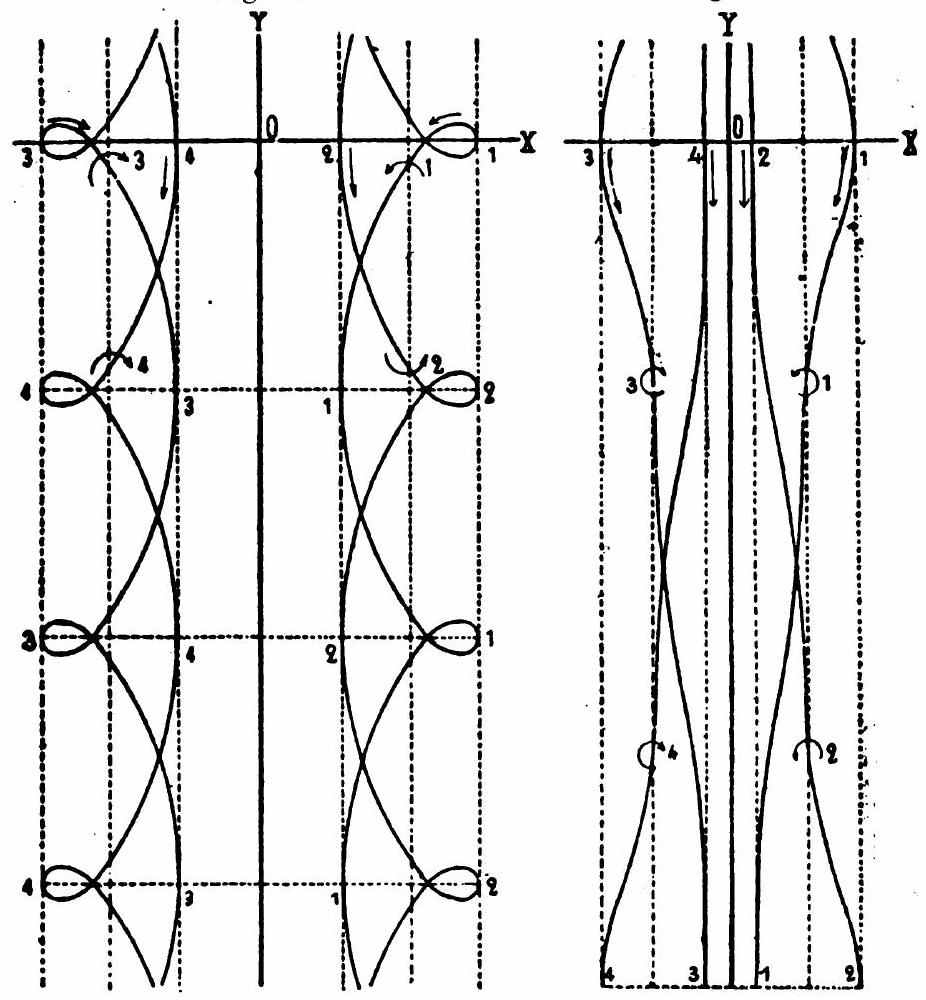}
  \caption{Gröbli's Figure 7 and 8. These are referred to as the left and right panels of Figure 7 in the English text.}
  \label{fig:78}
\end{figure}

The left and right panels of Figure~\ref{fig:78}, which correspond to the values $\kappa=\frac{1}{4}$ and $\kappa=\frac{4}{5}$,  give an approximate idea of the course of the movement.
\stepcounter{figure}

\section{Case II: $1>\lambda>0$}

When reducing the elliptic integrals, we find $\lambda$ is a modulus, so we set
\begin{equation} \label{16_33}
\lambda=\kappa,
\end{equation}
where $x$ must satisfy the conditions
\begin{equation} \label{16_34}
\frac{1}{\sqrt{1+\kappa}} \ge x \ge \sqrt{1-\kappa}.
\end{equation}
Let us set
\begin{equation} \label{16_35}
x=\sqrt{\frac{1-\kappa^{2} \sin^{2} \psi}{1+\kappa}},
\end{equation}
so the values $\psi=0$ and $\psi=\frac{\pi}{2}$ correspond to the limits of $x$, the integrand becomes
$$
\frac{\dd x}{\sqrt{\left(x-1+x^{2}\right)\left(1-(x+1) x^{2}\right)}}=-\frac{\dd \psi}{\sqrt{1-\kappa^{2} \sin^{2} \psi}}
$$
and from Eqs.~\eqref{14_13},~\eqref{14_17},~\eqref{14_18} etc. the following equations result
\begin{align}
t & =-\frac{2  \kappa^{2}}{1- \kappa^{2}} E( \kappa, \psi)+\frac{ \kappa}{1- \kappa}  \tan \psi \sqrt {1- \kappa^{2} \sin^{2} \psi} ;\label{15_36}\\
y_1 & =\frac{2}{1- \kappa^{2}} E( \kappa, \psi)-2 F( \kappa, \psi)-\left(\frac{\sqrt{1 - \kappa^{2} \sin^{2} \psi}}{1- \kappa}-\sqrt{1+ \kappa}\right)  \tan \psi ; \label{15_37}\\
 y_2 &= \frac{2}{1- \kappa^{2}} E( \kappa, \psi)-2 F( \kappa, \psi)-\left(\frac{\sqrt{1- \kappa^ {2} \sin^{2} \psi}}{1- \kappa}+\sqrt{1+ \kappa}\right)  \tan \psi ; \label{15_38} \\
 \dv{x}{t}& =- \kappa \sqrt{1+ \kappa} \frac{\sin \psi \cos^{3} \psi}{\left(1+ \kappa \sin^{2} \psi \right)^{2}}; \label{15_39}\\
 \dv{y_1}{t}& =-\frac{(1+ \kappa)^{2}- \kappa^{2} \cos^{4} \psi-(1+ \kappa)^{3 / 2} \sqrt{1- \kappa^{2} \sin^{2} \psi}}{ \kappa\left(1+ \kappa \sin^{2} \psi\right)^{2}} \label{15_40}; \\
 \dv{y_2}{t}& =-\frac{(1+ \kappa)^{2}- \kappa^{2} \cos^{4} \psi+(1+ \kappa)^{3 / 2} \sqrt{1- \kappa^{2} \sin^{2} \psi}}{ \kappa\left(1+ \kappa \sin^{2} \psi\right)^{2}}. \label{15_41}
\end{align}
The integration constants are determined so that the quantities $t$, $\psi$, $y_1$, and $y_2$ disappear simultaneously.

For $\psi=0$, it follows from these equations
$$
\begin{gathered}
t=0, x=\frac{1}{\sqrt{1+ \kappa}}, \quad y_1=0, \quad y_2=0 ,\\
\dv{x}{t}=0, \dv{y_1}{t}=-\frac{1+2  \kappa-(1+ \kappa)^{3 / 2}}{ \kappa}, \dv{y_2}{t}=-\frac{1+2  \kappa+(1+ \kappa)^{3 / 2}}{ \kappa},
\end{gathered}
$$
and for $\psi=\frac{\pi}{2}$
$$
\begin{aligned}
t & =\infty, \quad x=\sqrt{1- \kappa}, \quad y_1=-\infty, \quad y_2=-\infty , \\
\dv{x}{t} & =0, \dv{y_1}{t}=-\frac{1-\sqrt{1- \kappa}}{ \kappa}, \dv{y_{2 }}{t}=-\frac{1+\sqrt{1- \kappa}}{ \kappa} .
\end{aligned}
$$
The quantities $\dv{x}{t}$, $\dv{y_1}{t}$,  and $\dv{y_2}{t}$ are all consistently negative in the interval $0<\psi<\frac{\pi}{2}$. The correctness of this assertion for the first and third of these quantities can be seen directly by looking at the relevant equations. To prove that $\dv{y_1}{t}$ is negative, one has to show that
$$
(1+ \kappa)^{2}- \kappa^{2} \cos^{4} \psi >  (1+ \kappa)^{3/2} \sqrt{1- \kappa^{2} \sin^{ 2} \psi}.
$$
The left side of this inequality exceeds $1+2  \kappa$, and the right side is less than $(1+ \kappa)^{3 / 2}$, so that
$$
1+2  \kappa>(1+ \kappa)^{3/2}, 
$$
so our claim is proven. From the moment $t=0$ onwards, $x$, $y_1$, and $y_2$ continually decrease.

\begin{figure}[htbp] 
   \centering
   \includegraphics[width=0.45\textwidth]{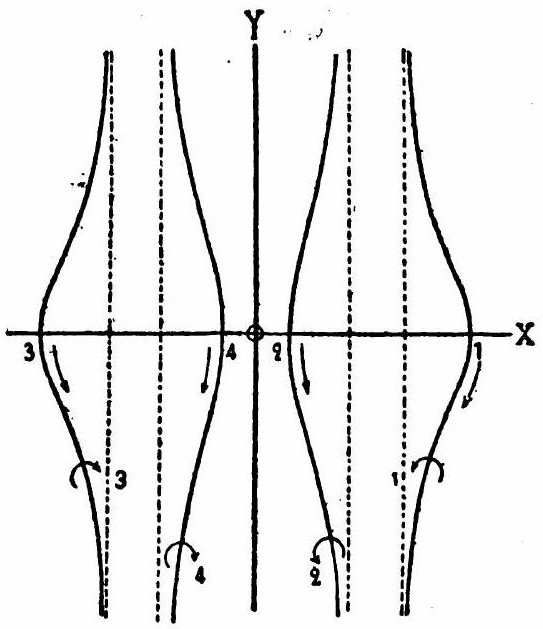} 
   \caption{Gröbli's figure 9.}
   \label{fig:9}
\end{figure}

At time $t=0$, the velocity points along the the negative $y$-axis and threads 1 and 2 lie on the $x$-axis with values
$$
x_1=1+\frac{1}{\sqrt{1+ \kappa}}, \quad x_2=1-\frac{1}{\sqrt{1+ \kappa}}.
$$
 Both $y_1$ and $y_2$ are constantly decreasing in Fig.~\ref{fig:9}; thread 2 leads thread 1 so that the distance between the threads from each other increases indefinitely. $x_1$ decreases, $x_2$ increases. The excursion in the direction of the $x$-axis is
$
=\frac{1}{\sqrt{1+ \kappa}}-\sqrt{1- \kappa} .
$
The straight lines
$$
 x_1=1+\sqrt{1- \kappa} \qand x_2=1-\sqrt{1- \kappa}
$$
are asymptotes of the curves described by the two threads. Curve 1 approaches its asymptote much more quickly than curve 2. Each of the curves has an inflection point for negative $y$. Figure~\ref{fig:9}, which corresponds to the value $x=\frac{24}{25}$, should give an idea of the course of the movement.

\section{Case III: $\lambda = 1$}
Setting
\begin{equation} \label{17_42}
x=\frac{\cos \psi}{\sqrt{2}}
\end{equation}
results in the following equations
\begin{align}
t & =\frac{1}{2} \sin \psi+\frac{\sin \psi}{\cos^{2} \psi}-\log \tan\left(\frac{\pi}{4} +\frac{\psi}{2}\right); \label{17_43}\\
y_1 & =\frac{1}{2} \sin \psi-\frac{\sin \psi}{\cos^{2} \psi}+\sqrt{2} \tan \psi-\log \tan\left(\frac{\pi}{4}+\frac{\psi}{2}\right); \label{17_44}\\
y_2 & =\frac{1}{2} \sin \psi-\frac{\sin \psi}{\cos^{2} \psi}-\sqrt{2} \tan \psi-\log \tan\left(\frac{\pi}{4}+\frac{\psi}{2}\right) \label{17_45}.
\end{align}
The remaining formulas are obtained from the corresponding ones in the previous paragraph by setting $x=1$.

\section{Case IV: $0 > \lambda>-1$}

In this case, $x$ must satisfy the conditions
$$
\frac{1}{\sqrt{1+\lambda}} \geqq x \geqq \sqrt{1-\lambda}.
$$
Since $x>1$, threads 1 and 2 lie on different sides of the symmetry axis. We set the size $-\lambda$ as the modulus of the elliptic integrals
\begin{equation} \label{18_46}
\lambda=-\kappa,
\end{equation}
so the conditions on $x$ become
\begin{equation} \label{18_47}
\frac{1}{\sqrt{1-\kappa}} \ge x \ge \sqrt{1+\kappa}
\end{equation}
above. These conditions differ from those of the case $1>\lambda>0$ only in that $\kappa$ is replaced by $-\kappa$. Therefore, the substitution
\begin{equation} \label{18_48}
x=\sqrt{\frac{1-\kappa^{2} \sin^{2} \psi}{1-\kappa}}
\end{equation}
can be used, and all formulas can be obtained directly from the previous ones by replacing $\kappa$ with $-\kappa$. In the equation that results in this way for $t$, positive values of $\psi$ correspond to negative values of $t$. To avoid this, we replace $\psi$ everywhere with $-\psi$; in other words, we define the root appearing in equations~\eqref{14_13} and~\eqref{14_17} with a negative sign. So we get the following equations
\begin{align}
t&=\frac{2 \kappa^{2}}{1-\kappa^{2}} E(\kappa, \psi)+\frac{\kappa}{1+\kappa} \tan{\psi} \sqrt{1-\kappa^ {2} \sin^{2} \psi}; \label{18_49}\\
y_1&=\frac{-2}{1-\kappa^{2}} E(\kappa, \psi)+2 F(\kappa, \psi)+\left(\frac{\sqrt{ 1-\kappa^{2} \sin^{2} \psi}}{1+\kappa}-\sqrt{1-\kappa}\right)\tan \psi; \label{18_50} \\
y_2&=\frac{-2}{1-\kappa^{2}} E(\kappa, \psi)+2 F(\kappa, \psi)+\left(\frac{\sqrt{ 1-\kappa^{2} \sin^{2} \psi}}{1+\kappa}+\sqrt{1-\kappa}\right)\tan \psi; \label{18_51} \\
\dv{x}{t}&=-\kappa \sqrt{1-\kappa} \frac{\sin \psi \cos^{3} \psi}{\left(1-\kappa \sin^{2} \psi \right)^{2}} ;\label{18_52}\\
\dv{y_1}{t}&=\frac{(1-\kappa)^{2}-\kappa^{2} \cos^{4} \psi-(1-\kappa)^{3 / 2} \sqrt{1-\kappa^{2} \sin^{2} \psi}}{\kappa\left(1-\kappa \sin^{2} \psi\right)^{2}}; \label{18_53}\\
\dv{y_2}{t}&=\frac{(1-\kappa)^{2}-\kappa^{2} \cos^{4} \psi+(1-\kappa)^{3 / 2 } \sqrt{1-\kappa^{2} \sin^{2} \psi}}{\kappa\left(1-\kappa \sin^{2} \psi\right)^{2}} . \label{18_54}
\end{align}
For $\psi=0$, this gives
$$
\begin{gathered}
t=0, x=\frac{1}{\sqrt{1-\kappa}}, y_1=0, \quad y_2=0, \\
\dv{x}{t}=0, \dv{y_1}{t}=\frac{1-2 \kappa-(1-\kappa)^{3 / 2}}{\kappa}, \dv {y_2}{t}=\frac{1-2 \kappa+(1-\kappa)^{3 / 2}}{\kappa},
\end{gathered}
$$
and for $\psi=\frac{\pi}{2}$,
$$
\begin{gathered}
t=\infty, x=\sqrt{1+\kappa}, \quad y_1=-\infty, \quad y_2=\infty \\
\dv{x}{t}=0, \quad \dv{y_1}{t}=\frac{1-\sqrt{1+\kappa}}{\kappa}, \quad \dv{y_ {2}}{t}=\frac{1+\sqrt{1+\kappa}}{\kappa} .
\end{gathered}
$$

\begin{figure}[htbp] 
   \centering
   \includegraphics[width=.5\textwidth]{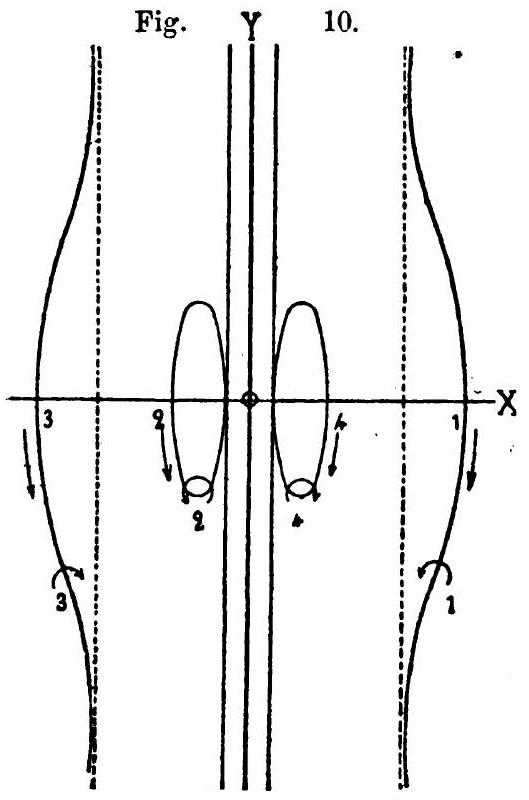} 
   \caption{Gröbli's Figure 10.}
   \label{fig:10}
\end{figure}

On the interval $0<\psi<\frac{\pi}{2}$,  $\dv{x}{t}$ is always negative. Since $\dv{y_1}{t}$ is negative at both the beginning and the end of this interval, the number of values for which it now vanishes must be even. In order for $\dv{y_1}{t}=0$, the equation
$$
(1-\kappa)^{2}-\kappa^{2} \cos^{4} \psi=(1-\kappa)^{3 / 2} \sqrt{1-\kappa^{2} \sin^{2} \psi}
$$
must be satisfied. The left side increases with $\psi$, and the right side decreases, so there can be at most one root. It follows that $\dv{y_1}{t}$ does not disappear at all, and $y_1$ continually decreases. For $\psi=\frac{\pi}{2}$ $\dv{y_2}{t}$ is positive, for $\psi=0$ it is positive or negative, depending
on whether
$$
x \gtrless \frac{\sqrt{5}-1}{2}.
$$
In the first case, $\dv{y_2}{t}$ is always positive; in the second, it has a single zero; the curve traversed by thread 2 has a self-crossing point, and, possibly, for $x=\frac{\sqrt{5}-1}{2}$, a corner. Curve 1 has an inflection point, which is also the value of $x$, while curve 2 only has an inflection point if $x<\frac{\sqrt{5}-1}{2}$. Figure~\ref{fig:10} corresponds to the value $x=\frac{4}{5}$.

\section{Case V: $-1>\lambda>-\infty$}

We set
\begin{equation} \label{19_55}
\lambda=-\frac{1}{\kappa}.
\end{equation}
Then $\kappa$ becomes the modulus of the elliptic integrals, and $x$ must meet the condition
\begin{equation} \label{19_56}
\sqrt{\frac{1+x}{x}}<x<\infty.
\end{equation}
By applying the substitution
\begin{equation} \label{19_57}
x=\sqrt{\frac{1+\kappa}{\kappa}} \frac{1}{\cos \psi},
\end{equation}
if the integration constants are determined appropriately, the following equations are obtained
\begin{align}
t &=\frac{-2}{\kappa\left(1-\kappa^{2}\right)} E(\kappa, \psi)+\frac{2}{\kappa} F(\kappa, \psi)+\frac{1}{\kappa}\left(\frac{1}{1-\kappa} \tan{\psi}-\frac{1}{1+\kappa} \cot \psi\right) \sqrt{1-\kappa^{2} \sin^{2} \psi}; \label{19_58} \\
y_1& =\frac{2 \kappa}{1-\kappa^{2}} E(\kappa, \psi)-\left(\frac{1}{1-\kappa} \tan{\psi}+\frac{1}{1+\kappa} \cot \psi-\frac{1}{\sqrt{\kappa(1+\kappa )} \sin \psi}\right) \sqrt{1-\kappa^{2} \sin^{2} \psi} ; \label{19_59} \\
y_2& =\frac{2 \kappa}{1-\kappa^{2}} E(\kappa, \psi)-\left(\frac{1}{1-\kappa} \tan{\psi}+\frac{1}{1+\kappa} \cot \psi+\frac{1}{\sqrt{\kappa(1+\kappa) } \sin \psi}\right) \sqrt{1-\kappa^{2} \sin^{2} \psi} ; \label{19_60} \\
\dv{x}{t}&=(1+\kappa) \sqrt{\kappa(1+\kappa)} \frac{\sin^{3} \psi \sqrt{1-\kappa^{2} \sin^{ 2} \psi}}{\left(1+\kappa \sin^{2} \psi\right)^{2}} ; \label{15_61} \\
\dv{y_1}{t}&=-\frac{-\kappa+2 \kappa \sin^{2} \psi+\kappa^{2}(2+\kappa) \sin^{4 } \psi+\sqrt{\kappa(1+\kappa)} \cos^{3} \psi}{\left(1+\kappa \sin^{2} \psi\right)^{2}} ; \label{15_62} \\
\dv{y_2}{t}& =-\frac{-\kappa+2 \kappa \sin^{2} \psi+\kappa^{2}(2+\kappa) \sin^{4 } \psi-\sqrt{\kappa(1+\kappa)} \cos^{3} \psi}{\left(1+\kappa \sin^{2} \psi\right)^{2}} ; \label{15_63}
\end{align}

For $\psi=0$, this is
$$
\begin{aligned}
&t=-\infty, x=\sqrt{\frac{1+\kappa}{\kappa}}, y_1=\infty, y_2=-\infty, \\
&\dv{x}{t}=0, \dv{y_1}{t}=\kappa-\sqrt{\kappa(1+\kappa)}, \dv{y_2}{t}= \kappa+\sqrt{\kappa(1+\kappa)},
\end{aligned}
$$
and for $\psi=\frac{\pi}{2}$
$$
\begin{aligned}
& t=\infty, \quad x=\infty, \quad y_1=-\infty, \quad y_2=-\infty, \\
& \dv{x}{t}=\sqrt{\kappa(1+\kappa)}, \quad \dv{y_1}{t}=\dv{y_2}{t}=- \kappa.
\end{aligned}
$$

\begin{figure}[htbp] 
   \centering
   \includegraphics[width=.5\textwidth]{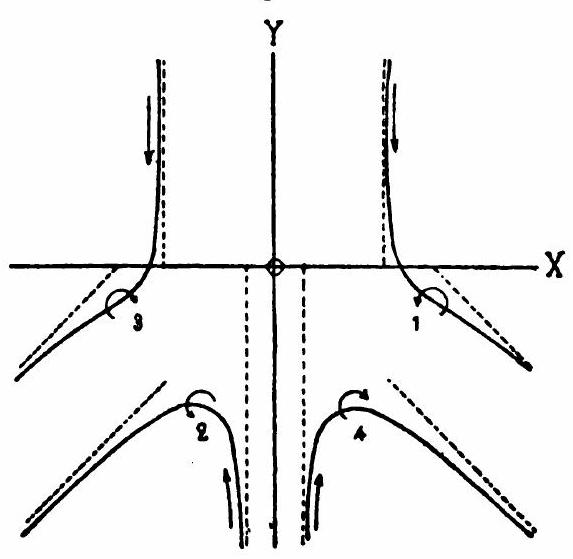} 
   \caption{Gröbli's Figure 11}
   \label{fig:11}
\end{figure}

The coordinate $y_1$ constantly decreases, while $y_2$ first increases, then decreases. The straight lines
\begin{equation} \label{19_64}
x_1=1+\sqrt{\frac{1+\kappa}{\kappa}} \qand 
x_2=1-\sqrt{\frac{1+\kappa}{\kappa}}
\end{equation}
are asymptotes of the trajectories of threads 1 and 2. Each of the curves has a second asymptote, which is defined by
\begin{equation} \label{19_65}
\begin{split}
y_1&=-\sqrt{\frac{\kappa}{1+\kappa}} x_1+\frac{2 \kappa}{1-\kappa^{2}} E+\frac{1}{\sqrt {\kappa(1-\kappa)}} \\
y_2&=\sqrt{\frac{\kappa}{1+\kappa}} x_2+\frac{2 \kappa}{1-\kappa^{2}} E-\frac{1}{\sqrt {\kappa(1-\kappa)}} .
\end{split}
\end{equation}
The curve traversed by thread 1 has an inflection point. Figure~\ref{fig:11} corresponds to the value $\kappa=\frac{1}{2}$.

\section{Case VI: $\lambda=-1$}

Setting
\begin{equation} \label{20_66}
x=\frac{\sqrt{2}}{\cos \psi}
\end{equation}
results in the following equations if the integration constants are determined appropriately
\begin{align}
t & =\frac{\sin \psi}{\cos^{2} \psi}-\frac{1}{2 \sin \psi}+\log \tan\left(\frac{\pi}{4} +\frac{\psi}{2}\right); \label{20_67}\\
y_1 & =-\frac{\sin \psi}{\cos^{2} \psi}-\frac{1}{2 \sin \psi}+\frac{1}{\sqrt{2} } \cot \psi+\log \tan\left(\frac{\pi}{4}+\frac{\psi}{2}\right); \label{20_68} \\
y_2 & =-\frac{\sin \psi}{\cos^{2} \psi}-\frac{1}{2 \sin \psi}-\frac{1}{\sqrt{2} } \cot \psi+\log \tan\left(\frac{\pi}{4}+\frac{\psi}{2}\right). \label{20_69}
\end{align}
The remaining formulas are obtained for $\kappa=1$ from the corresponding ones in the previous paragraph.

\section{On the motion of $2 n$ vortices, assuming $n$ planes of symmetry}

For the movement in the $xy$ plane, there are $n$ symmetry axes, which all have to pass through the same point and decompose the entire plane into $2 n$ congruent angular spaces. There is a thread in each of these. We place the intersection of the symmetry axes at the origin and make one of the symmetry axes the $x$-axis. From here, going around the starting point in a positive sense, the threads $1, 2,\ldots, 2n$ should follow one another in sequence. The necessary and sufficient conditions which must be fulfilled for the assumed movement to be possible are
\begin{equation} \label{21_1}
m_1=-m_2=m_3=-m_{4}=\cdots =m_{2 n-1}=-m_{2 n} .
\end{equation}
We assume $m_1$ to be positive, and by appropriate scaling of time, we can then set the collective value of these quantities equal to $2 \pi$.

Let $\varrho_1, \vartheta_1; \varrho_2, \vartheta_2 ; \ldots; \varrho_{2 n}, \vartheta_{2 n}$ be the polar coordinates of the vortex threads and assume that
\begin{equation} \label{21_2}
\varrho_1=\varrho_2=\cdots =\varrho_{2 n},
\end{equation}
and
\begin{equation} \label{21_3}
\begin{aligned}
\vartheta_2& =\frac{2 \pi}{n}-\vartheta_1, & \vartheta_3&=\frac{2 \pi}{n}+\vartheta_1, \\
\vartheta_{4}&=\frac{4 \pi}{n}-\vartheta_1, & \vartheta_{5}&=\frac{4 \pi}{n}+\vartheta_1,  \\
\cdots & &\cdots& \\
\vartheta_{2 n}& =2 \pi-\vartheta_1, & \vartheta_{2 n-1}&=\frac{(2 n-2) \pi}{n}+\vartheta_1 .
\end{aligned}
\end{equation}

To determine the movement of the first thread, we use the equations
$$
m_1 \varrho_1 \dv{\varrho_1}{t}=\frac{\partial P}{\partial \vartheta_1}, \quad m_1 \varrho_{1} \dv{\vartheta_1}{t}=-\frac{\partial P}{\partial \varrho_1} .
$$
Using 
$$
P=-\frac{1}{\pi} \sum m_1 m_2 \log \varrho_{12}
$$
and Eq.~\eqref{21_2}, we find
$$
\frac{\partial P}{\partial \vartheta_1}=-\frac{m_1}{\pi} \sum m_2 \cot \frac{\vartheta_1-\vartheta_{ 2}}{2} .
$$
The sum is to be understood so that the indices $3, 4 \ldots, 2 n$ are placed one after the other in place of index 2. This equation can be written, taking Eqs.~\eqref{21_1} and~\eqref{21_3} into account, as
$$
\begin{aligned}
\frac{\partial P}{\partial \vartheta_1}  = & \phantom{+}2 \pi\left\{\cot\left(\vartheta_1-\frac{\pi}{n}\right)+\cot\left(\vartheta_1-\frac{2 \pi}{n}\right)+\cdots+\cot\left(\vartheta_1-\pi\right)\right\} \\
& +2 \pi\left\{\cot \frac{\pi}{n}+\cot \frac{2 \pi}{n}+\cdots +\cot \frac{(n-1) \pi}{n}\right\} .
\end{aligned}
$$
The expression in the second bracket disappears because the terms cancel each other away from the two ends in pairs, and the middle term, which is present when $n$ is even, disappears by itself. The row in the first bracket is equal to $n \cot n \vartheta_1$, and therefore we find
$$
\frac{\partial P}{\partial \vartheta_1}=2 n \pi \cot {n \vartheta_1} .
$$
From the above equation for $P$, it also follows that
$$
\frac{\partial P}{\partial \varrho_1}=\frac{2 \pi}{\varrho_1},
$$
and these yield the following equations
\begin{equation} \label{21_4}
\varrho_1 \dv{\varrho_1}{t}=n \cot n \vartheta_1, \quad 
\varrho_1^{2} \dv{\vartheta_1} {t}=-1 .
\end{equation}
From these, it follows by eliminating $t$,
$$
\frac{\dd \varrho_1}{\varrho_1}=-n \cot n \vartheta_1 \cdot \dd \vartheta_1
$$
and from this, through integration,
\begin{equation} \label{21_5}
\varrho_1 \sin n \vartheta_1=1,
\end{equation}
if, as is permitted, we assign a special value to the constant of integration.

This equation represents the trajectory of thread 1 if we let $\vartheta_1$ go from 0 to $\frac{\pi}{n}$. If we give $\vartheta_1$ all values from 0 to $2 \pi$, then from Eq.~\eqref{21_5}, we get not only the path of the first thread, but also the paths of all vortex threads if $n$ is even, and those of odd $n$ paths of threads $1,3,5, \ldots, 2 n-1$. The curves described by the threads $2,4, \ldots 2 n$ are contained in the equation
$$
\varrho_2 \sin n \vartheta_2=-1
$$
We can, therefore, generally say that the equation
\begin{equation} \label{21_6}
\varrho^{2} \sin^{2} n \vartheta=1
\end{equation}
represents the paths of all vortex threads. This equation represents a curve of order $2 n$, which consists of $2 n$ congruent branches, and, for an odd $n$, decomposes into the two curves of $n$th order
$$
\varrho \sin n \vartheta = \pm 1.
$$
 The straight lines that bisect the angles between the axes of symmetry of the movement are the axes of symmetry of the curve. It follows from Eq.~\eqref{21_6}
$$
y^{2}=\frac{\sin^{2} \vartheta}{\sin^{2} n \vartheta},
$$
and from this, for infinitesimally small $\vartheta$
$$
y^{2}=\frac{1}{n^{2}}.
$$
Here, $\varrho$ becomes infinitely large, and therefore, the lines parallel to the symmetry axes of the movement, at a distance $\frac{1}{n}$ from them, are asymptotes.

From Eq.~\eqref{21_5} and the second equation of~\eqref{21_4}, we find, by eliminating $\varrho_1$ and squaring
\begin{equation} \label{21_7}
\cot n \vartheta_1=n t
\end{equation}
and then from Eq.~\eqref{21_5},
\begin{equation}\label{21_8}
\varrho_1^{2}=1+n^{2} t^{2}.
\end{equation}
The time is measured from the moment $\varrho_1=1$ and $\vartheta_1=\frac{\pi}{2 n}$. From here, $\varrho_1$ increases and $\vartheta_1$ decreases.

The resulting equation for the speed is
\begin{equation} \label{21_9}
 w_1^2=\frac{1+n^4 t^2}{1+n^2 t^2} = \frac{1-n^2+n^2 \varrho_1^2}{\varrho_1^2} .
\end{equation}
The velocity reaches a minimum value $w_1=1$ at $t=0$, then increases steadily as it approaches the limit $n$ as $t \to \infty$.

\begin{figure}[htbp] 
   \centering
   \includegraphics[width=0.5\textwidth]{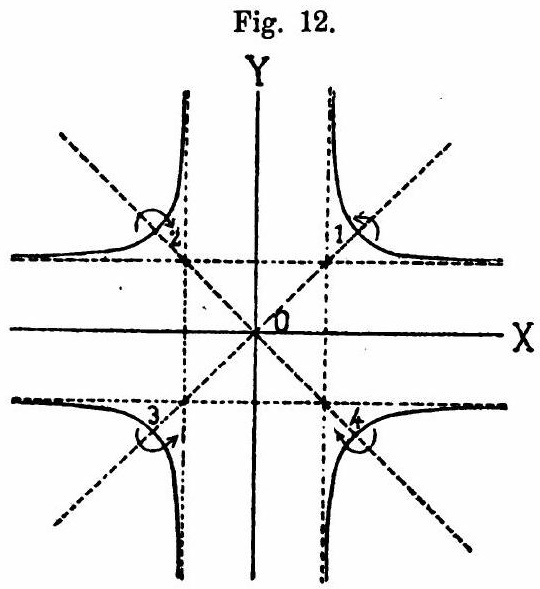} 
   \caption{Gröbli's Figure 12.}
   \label{fig:12}
\end{figure}

Figure~\ref{fig:12} corresponds to the assumption $n=2$.

\end{document}